\documentclass[pre,preprint,showpacs,preprintnumbers]{revtex4}

\usepackage{graphicx}
\usepackage{latexsym}
\usepackage{amsmath}
\usepackage{amsfonts}
\usepackage{bm}

\newcommand{\la}{\left<}
\newcommand{\ra}{\right>}
\newcommand{\Rend}{\ensuremath{R_\text{e}}}
\newcommand{\Rgyr}{\ensuremath{R_\text{g}}}
\newcommand{\be}{\ensuremath{b_\text{e}}}
\newcommand{\bg}{\ensuremath{b_\text{g}}}
\newcommand{\ce}{\ensuremath{c_\text{e}}}
\newcommand{\cp}{\ensuremath{c_p}}
\newcommand{\cP}{\ensuremath{c_\text{P}}}
\newcommand{\Acal}{\mbox{$\cal A$}}
\newcommand{\alphap}{\ensuremath{\alpha_p}}
\newcommand{\Kp}{\ensuremath{K_p}}

\newcommand{\Ustar}{u}
\newcommand{\veff}{\tilde{v}}
\newcommand{\Gi}{\ensuremath{G_\text{z}}}

\newcommand{\taue}{\ensuremath{\tau_\text{e}}}
\newcommand{\Ds}{\ensuremath{D_\text{s}}}
\newcommand{\Dc}{\ensuremath{D_\text{c}}}
\newcommand{\Ne}{\ensuremath{N_\text{e}}}

\newcommand{\nch}{\ensuremath{n_\text{ch}}}
\newcommand{\nmon}{\ensuremath{n_\text{mon}}}
\newcommand{\lp}{\ensuremath{l_\text{p}}}
\newcommand{\kB}{\ensuremath{k_\text{B}}}

\begin{document}

\title{Intramolecular long-range correlations in polymer melts:\\ 
The segmental size distribution and its moments}

\author{J.P.~Wittmer}
\email{jwittmer@ics.u-strasbg.fr}
\affiliation{Institut Charles Sadron, CNRS, 23 rue du Loess, 67037 Strasbourg C\'edex, France}
\author{P.~Beckrich}
\affiliation{Institut Charles Sadron, CNRS, 23 rue du Loess, 67037 Strasbourg C\'edex, France}
\author{H.~Meyer}
\affiliation{Institut Charles Sadron, CNRS, 23 rue du Loess, 67037 Strasbourg C\'edex, France}
\author{A.~Cavallo}
\affiliation{Institut Charles Sadron, CNRS, 23 rue du Loess, 67037 Strasbourg C\'edex, France}
\author{A.~Johner}
\affiliation{Institut Charles Sadron, CNRS, 23 rue du Loess, 67037 Strasbourg C\'edex, France}
\author{J.~Baschnagel}
\homepage{http://www-ics.u-strasbg.fr/~etsp/welcome.php}
\affiliation{Institut Charles Sadron, CNRS, 23 rue du Loess, 67037 Strasbourg C\'edex, France}

\date{\today}

\begin{abstract}
Presenting theoretical arguments and numerical results we demonstrate long-range intrachain 
correlations in concentrated solutions and melts of long flexible polymers which cause a
systematic swelling of short chain segments. They can be traced 
back to the incompressibility of the melt leading to an effective repulsion
$\Ustar(s) \approx s/\rho R^3(s) \approx \ce/\sqrt{s}$ when connecting two segments together 
where $s$ denotes the curvilinear length of a segment, 
$R(s)$ its typical size, 
$\ce \approx 1/\rho \be^3$ the ``swelling coefficient",
$\be$ the effective bond length and $\rho$ the monomer density.
The relative deviation of the segmental size distribution from the ideal Gaussian chain behavior
is found to be proportional to $\Ustar(s)$. The analysis of different moments of this distribution 
allows for a precise determination of the effective bond length $\be$ and 
the swelling coefficient $\ce$ of asymptotically long chains.
At striking variance to the short-range decay suggested by Flory's ideality hypothesis
the bond-bond correlation function of two bonds separated by $s$ monomers along the chain 
is found to decay algebraically as $1/s^{3/2}$.
Effects of finite chain length are considered briefly.
\end{abstract}

\maketitle

\section{Flory's ideality hypothesis revisited\label{sec_intr}}

\paragraph*{A cornerstone of polymer physics.}
Polymer melts are dense disordered systems consisting of macromolecular chains 
\cite{RubinsteinBook}. Theories that predict properties of chains in a melt or 
concentrated solutions generally start from the ``Flory ideality hypothesis" 
formulated already in the 1940s by Flory \cite{Flory45,Flory49,FloryBook}.
This cornerstone of polymer physics states that chain conformations correspond 
to ``ideal" random walks on length scales much larger than the monomer diameter 
\cite{FloryBook,DegennesBook,DoiEdwardsBook,RubinsteinBook}.
The commonly accepted justification of this mean-field result is that {\em intra}chain 
and {\em inter}chain excluded volume forces compensate each other if many chains  
strongly overlap which is the case for three-dimensional melts \cite{DegennesBook}. 
Since these systems are essentially incompressible, density fluctuations are known to 
be small. Hence, all correlations are supposed to be short-ranged as has been systematically 
discussed first by Edwards who developed the essential statistical mechanical tools 
\cite{E65,E66,E75,ME82,DoiEdwardsBook} also used in this paper.
One immediate consequence of Flory's hypothesis is that the mean-squared 
size of chain segments of curvilinear length $s=m-n$ (with $1 \le n < m  < N$) 
should scale as $\Rend^2(s) \equiv \la {\bf r}^2 \ra = {\be^2s}$
if the two monomers $n$ and $m$ on the same chain are sufficiently separated
along the chain backbone, and local correlations may be neglected ($1 \ll s$). 
For the total chain ($s=N-1 \gg 1$) this implies obviously that 
$\Rend^2(N-1) = \be^2 (N-1) \approx \be^2 N$. 
Here, $N$ denotes the number of monomers per chain, 
${\bf r}$ the end-to-end vector of the segment,
$r = ||{\bf r}||$ its length and 
$\be$ the ``effective bond length" of asymptotically long chains \cite{DoiEdwardsBook}.
(See Fig.~\ref{fig_diagram} for an illustration of some notations used in this paper.)
For the $2p$-th moment ($p=0,1,2,\ldots$) of the segmental size distribution $G(r,s)$ 
in three dimensions one may write more generally
\begin{equation}
\Kp(s) \equiv 1 - \frac{6^p p!}{(2p+1)!} \frac{\la {\bf r}^{2p} \ra}{(\be^{2}s)^p} = 0
\label{eq_Kpdef}
\end{equation}
which is, obviously, consistent with a Gaussian segmental size distribution
\begin{equation}
G_0(r,s) = \left(\frac{3}{2\pi s \be^2}\right)^{3/2} \exp\left(- 
\frac{3}{2} \frac{r^2}{\be^2 s}\right).
\label{eq_G0rs}
\end{equation}
Both equations are expected to hold as long as the moment is not too high for a 
given segment length and the finite-extensibility of the polymer strand remains 
irrelevant \cite{DoiEdwardsBook}.
\paragraph*{Deviations caused by the segmental correlation hole effect.}
Recently, Flory's hypothesis has been challenged both theoretically
\cite{SchaferBook,ANS03,ANS05a,BJSOBW07,BeckrichThesis} and numerically for three-dimensional 
solutions \cite{SMK00,Auhl03,WMBJOMMS04,WBJSOMB07,WBCHCKM07} 
and ultrathin films \cite{CMWJB05,MKCWB07}.
These studies suggest that intra- and interchain excluded volume forces do not fully
compensate each other on intermediate length scales, leading to long-range intrachain
correlations.
The general physical idea behind these correlations is related 
to the ``segmental correlation hole" of a typical chain segment \cite{WBJSOMB07}.
As sketched in Fig.~\ref{fig_corehole}, this induces an effective repulsive interaction 
when bringing two segments together, and {\em swells} (to some extent) the chains
causing, hence, a systematic violation of Eq.~(\ref{eq_Kpdef}).
Elaborating and clarifying various points already presented briefly elsewhere 
\cite{WMBJOMMS04,WBJSOMB07,WBCHCKM07}, we focus here on melts of long and flexible polymers. 
Using two well-studied coarse-grained polymer models \cite{BWM04} various intrachain properties 
are investigated numerically as functions of $s$ and compared with predictions from first-order 
perturbation theory. 
(For a discussion of intrachain correlations in reciprocal space see 
Refs.~\cite{WBJSOMB07,BJSOBW07,BeckrichThesis}.)
\paragraph*{Central results tested in this study.}
The key claim verified here concerns the deviation $\delta G(r,s) = G(r,s) - G_0(r,s)$ of
the segmental size distribution $G(r,s)$ from Gaussianity, Eq.~(\ref{eq_G0rs}),
for asymptotically long chains ($N\rightarrow \infty$) in the ``scale-free regime" 
($1 \ll s \ll N$). We show that the relative deviation divided by $\ce/\sqrt{s}$
scales as a function $f(n)$ of $n=r/\be \sqrt{s}$:
\begin{equation}
\frac{\delta G(r,s)/G_0(r,s)}{\ce/\sqrt{s}} =
f(n) = \sqrt{\frac{3\pi}{32}} \left( -\frac{2}{n} + 9 n - \frac{9}{2} n^3 \right).
\label{eq_dGrsbe}
\end{equation}
%
As we shall see, this scaling holds indeed for sufficiently 
large segment size $r$ and curvilinear length $s$.
The indicated ``swelling coefficient" $\ce$ has been predicted analytically, 
\begin{equation}
\ce = \frac{\sqrt{24/\pi^3}}{\rho \be^3}
\label{eq_cpce}
\end{equation}
($\rho$ being the monomer number density), where we shall argue that the bond 
length of the Gaussian reference chain of the perturbation calculation must 
be {\em renormalized} to the effective bond length $\be$. 
Accepting Eq.~(\ref{eq_dGrsbe}) the swelling of the segment size is readily 
obtained by computing $\la {\bf r}^{2p} \ra = 4\pi \int dr \ r^{2+2p} G(r,s)$.
For the $2p$-th moment this yields
\begin{equation}
\Kp(s) =  \frac{3 (2^p p! p)^2}{2 (2p+1)!} \ \frac{\cp}{\sqrt{s}}.
\label{eq_Rs_key}
\end{equation}
For instance, for the second moment ($p=1$) this reduces to 
$K_1(s) = 1 - \Rend(s)^2/\be^2s = c_1/\sqrt{s} \approx \ce/\sqrt{s}$.
We have replaced in Eq.~(\ref{eq_Rs_key}) the theoretically expected 
swelling coefficient $\ce$ by empirically determined coefficients $\cp$. 
It will be shown, however, that $\cp/\ce$ is close to unity for all moments.  
Effectively, this reduces Eq.~(\ref{eq_Rs_key}) to an efficient one-parameter 
extrapolation formula for the effective bond length $\be$ of asymptotically
long chains albeit empirical and theoretical swelling coefficients may slightly differ.
While we show how $\be$ may be {\em fitted}, no attempt is made to {\em predict} 
it from the operational model parameters and other measured properties
such as the microscopic structure or the bulk compression modulus 
\cite{E75,ME82,DoiEdwardsBook}.
\paragraph*{Outline.}
We begin our discussion by sketching the central theoretical ideas in 
Sec.~\ref{sec_theo}.
There we will give a simple scaling argument and outline very briefly some elements 
of the standard perturbation calculations we have performed to derive them 
(Sec.~\ref{sub_theopert}). 
%
Details of the analytical treatment are relegated to Appendix~\ref{app_pert}.
The numerical models and algorithms allowing the computation of dense melts containing 
the large chain lengths needed for a clear-cut test are presented in Sec.~\ref{sec_simu}.
Our computational results are given in Sec.~\ref{sec_resu}.
%
While focusing on long chains in dense melts, we explain also briefly  
effects of finite chain size.
%
The general background of this work and possible consequences for other problems of 
polymer science are discussed in the final Sec.~\ref{sec_conc}.

\section{Physical idea and sketch of the perturbation calculation\label{sec_theo}}

\subsection{Scaling arguments\label{sub_theoscal}}

\paragraph*{Incompressibility and correlation of composition fluctuations.}
Polymer melts are essentially incompressible on length scales large compared
to the monomer diameter, and the density $\rho$ of {\em all} monomers does not 
fluctuate. On the other hand, composition fluctuations of labeled chains or 
subchains may certainly occur, however, subject to the total density constraint.
Composition fluctuations are therefore coupled and segments feel an {\em entropic} 
penalty when their distance becomes comparable to their size \cite{ANS03,WBJSOMB07}.  
As sketched in Fig.~\ref{fig_corehole}(a), we consider two {\em independent} 
test chains of length $s$ in a melt of very long chains ($N \rightarrow \infty$). 
If $s$ is sufficiently large, their typical size, $R(s) \approx \be \sqrt{s}$, 
is set by the effective bond length $\be$ of the surrounding melt 
(taking apart finite chain-size effects). The test chains interact with each
other directly and through the density fluctuations of the surrounding melt.
The scaling of their effective interaction may be obtained from the potential 
of mean force $U(r,s) \equiv - \ln(g(r,s)/g(\infty,s))$ where $g(r)$ is the 
probability to find the second chain at a distance $r$ assuming the first 
segment at the origin ($r=0$). 
Since the correlation hole is shallow for large $s$, expansion leads to 
$U(r,s) \approx 1 - g(r,s)/g(\infty,s) \approx c(r,s)/\rho$ with $c(r,s)$ 
being the density distribution of a test chain around its center of mass. 
This distribution scales as $c(r\approx 0,s) \approx s/R(s)^d$ 
close to the center of mass ($d$ being the dimension of space) and 
decays rapidly at distances of order $R(s)$ \cite{DegennesBook}. 
Hence, the interaction strength at $r/R(s) \ll 1$ is set by 
$\Ustar(s) \equiv U(0,s) \equiv c(0,s)/\rho \approx s/\rho R(s)^d \sim s^{1-d/2}$ 
\cite{ANS03,WBJSOMB07}.  
Interestingly, $\Ustar(s)$ does not depend explicitly on the bulk compression modulus $v$. 
It is dimensionless and independent of the definition of the monomer unit, i.e. it does not 
change if $\lambda$ monomers are regrouped to form an effective monomer 
($\rho\to\rho/\lambda$, $s\to s/\lambda$) while keeping the segment size $R$ fixed.
\paragraph*{Connectivity and swelling.}
To connect both test chains to form a chain of length $2s$ the effective energy 
$\Ustar(s)$ has to be paid and this repulsion will push the half-segments apart.
We consider next a segment of length $s$ in the middle of a very long chain.
All interactions between the test segment and the rest of the chain are first
switched off but we keep all other interactions, especially within the segment 
and between the segment monomers and monomers of surrounding chains. 
The typical size $R(s)$ of the test segment remains essentially unchanged 
from the size of an independent chain of same strand length. 
If we now switch on the interactions between the segment and monomers on 
{\em adjacent} segments of same length $s$, 
this corresponds to an effective interaction of order $\Ustar(s)$ as before.
(The effect of switching on the interaction to all other monomers of the chain 
is inessential at scaling level, since these other monomers are more distant.)
Since this repels the respective segments from each other, the corresponding subchain 
is swollen compared to a Gaussian chain of non-interacting segments.
It is this effect we want to characterize. 

\paragraph*{Perturbation approach in three dimensions.}
In the following we will exclusively consider chain segments $s$ which are much larger 
than the number of monomers $g \equiv 1/v \rho$ contained in a blob \cite{DegennesBook}, 
i.e. we will look on a scale where incompressibility matters. (The number $g$ is also
sometimes called ``dimensionless compressibility" \cite{BJSOBW07}.) 
Interestingly, when taken at $s=g$ the interaction strength takes the value
\begin{equation}
\Ustar(s=g) \approx \frac{g}{\rho \be^d g^{d/2}} 
= \frac{(v\rho)^{d/2-1}}{\rho \be^d}
\approx \Gi
\label{eq_UstarGi}
\end{equation}
with $\Gi$ being the standard Ginzburg parameter used for the perturbation calculation of 
strongly interacting polymers \cite{DoiEdwardsBook}. Hence, the segmental correlation hole 
potential $\Ustar(s) \approx \Gi (g/s)^{d/2-1} \ll \Gi$ for $d > 2$ and $s \gg g$.
Although for real polymer melts as for computational systems large values of $\Gi \approx 1$ 
may sometimes be found, 
$\Ustar(s) \sim 1/\sqrt{s}$ decreases rapidly with $s$ in three dimensions, 
as illustrated in Fig.~\ref{fig_corehole}(b), and standard perturbation calculations 
can be successfully performed. 

As sketched in the next paragraph these calculations yield quantities $K[\Ustar]$ 
which are defined such that they vanish ($K[\Ustar=0]=0$) if the perturbation 
potential $\Ustar(s)$ is switched off and are then shown to scale, to leading order, 
linearly with $\Ustar$. 
For instance, for the quantity $\Kp(s)$, defined in Eq.~(\ref{eq_Kpdef}),
characterizing the deviation of the chain segment size from Flory's hypothesis
one thus expects the scaling
\begin{equation}
\Kp[\Ustar(s)] \approx + \Ustar(s) \approx + \frac{s}{\rho R^d(s)}.
\label{eq_RsUs}
\end{equation}
The $+$-sign indicated marks the fact that the prefactor has to be positive to be consistent 
with the expected swelling of the chains. Consequently, the typical segment size,
$R(s)/\be \sqrt{s} \approx 1 - \Ustar(s)$, must approach the asymptotic limit for large $s$ 
from below.
For three dimensional solutions Eq.~(\ref{eq_RsUs}) implies that $\Kp(s)$ should 
vanish rapidly as $1/(\rho \be^3 \sqrt{s})$.
(This is different in thin films where $\Ustar(s) \approx \Gi$ decays only 
logarithmically \cite{ANS03} as may be seen from Eq.~(\ref{eq_Rs_slit}) given below.)
Taking apart the prefactors --- which require a full calculation --- 
this corresponds exactly to Eq.~(\ref{eq_Rs_key}) with a
swelling coefficient $\ce \approx \cp \approx 1/\rho \be^3$
in agreement with Eq.~(\ref{eq_cpce}).
Note also that the predicted deviations are inversely proportional to $\be^3$, i.e. 
the more flexible the chains, the more pronounced the effect.
Similar relations $K[\Ustar] \sim \Ustar$ may also be formulated for other quantities
and will be tested numerically in Sec.~\ref{sec_resu}. There, we will also check that 
the linear order is sufficient. 

\subsection{Perturbation calculation}
\label{sub_theopert}

\paragraph*{Generalities.}
Before delving more into our computational results we summarize here how 
Eqs.~(\ref{eq_dGrsbe}-\ref{eq_Rs_key}) and related relations have been obtained 
using standard one-loop perturbation calculation.
The general task is to determine
$\la \Acal \ra \approx \la \Acal \ra_0 (1+ \la U \ra_0)-\la \Acal U \ra_0$
for measurable quantities \Acal \ such as the squared distance between two 
monomers $n$ and $m$ on the same chain, $\Acal = {\bf r}_{nm}^2$. 
Here, $\la ... \ra_0$ denotes the average over the distribution function
of the unperturbed ideal chain of bond length $b$
and $U = \int_0^N dk \, \int_0^k dl \, \veff(r_{kl})$
the effective perturbation potential.
We discuss first the general results in the scale free regime ($1 \ll s \ll N$),
argue then that $b$ should be renormalized to the effective bond length $\be$
and sketch finally the calculation of finite chain-size effects.

\begin{figure}[tb]
\includegraphics*[width=0.6\textwidth]{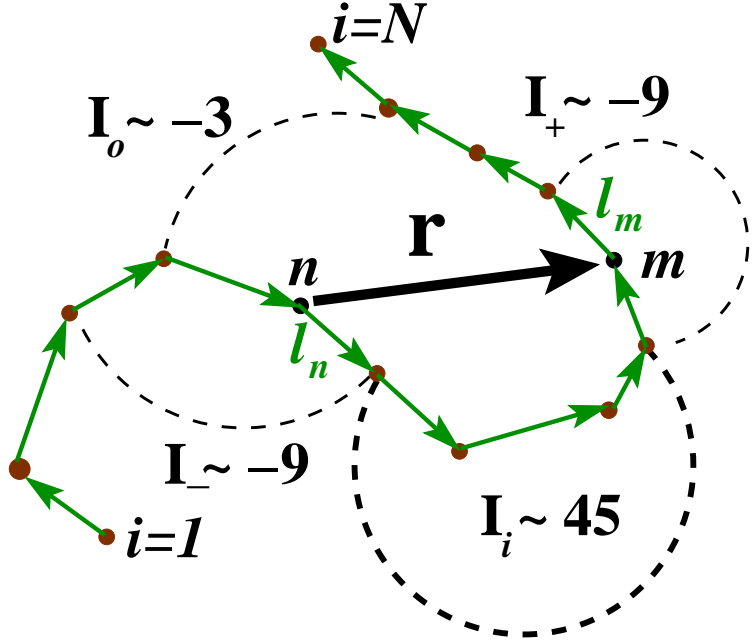}
\caption{(Color online)
Sketch of a polymer chain of length $N$ in a dense melt in $d=3$ dimensions.  
As notations we use ${\bf r}_i$ for the position vector of a monomer $i$,
${\bf l}_i={\bf r}_{i+1}-{\bf r}_{i}$ for its bond vector, 
${\bf r}={\bf r}_m-{\bf r}_n$ for the end-to-end vector of the chain segment 
between the monomers $n$ and $m=n+s$
and $r = || {\bf r} ||$ for its length. 
Segment properties, such as the $2p$-th moments $\la {\bf r}^{2p} \ra$, 
are averaged over all possible pairs of monomers $(n,m)$ of a chain
and over all chains. 
The second moment ($p=1$) is denoted $\Rend(s) = \la {\bf r}^2 \ra^{1/2}$,
the total chain end-to-end distance is $\Rend(s=N-1)$.
The dashed lines show the relevant graphs of the analytical perturbation calculation
outlined in Sec.~\ref{sub_theopert}.
The numerical factors indicate for infinite chains (without chain end effects) 
the relative weights contributing to the $1/\sqrt{s}$-swelling of $\Rend(s)$
indicated in Eq.~(\ref{eq_R2pert}).
\label{fig_diagram}
}
\end{figure}

\begin{figure}[tb]
\includegraphics*[width=0.9\textwidth]{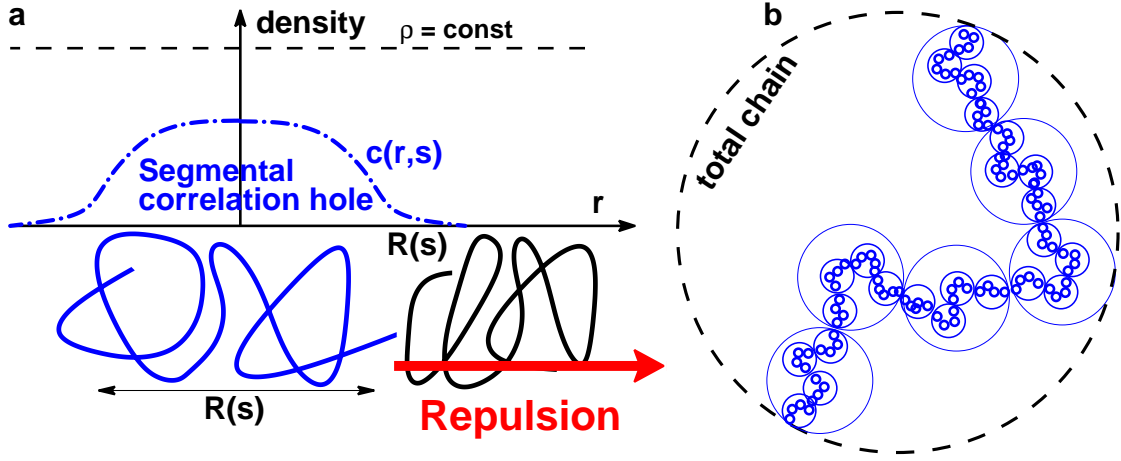}
\caption{(Color online)
\label{fig_corehole}
Role of incompressibility and chain connectivity in dense polymer solutions and melts.
{\bf (a)}
Sketch of the segmental correlation hole of a marked chain segment of curvilinear length $s$.
%
Density fluctuations of chain segments must be correlated, since the
total density fluctuations (dashed line) are small.
Consequently, a second chain segment 
feels an {\em entropic} repulsion when both correlation holes start to overlap.
{\bf (b)}
Self-similar pattern of nested segmental correlation holes of decreasing strength
$\Ustar(s) \approx s/\rho R(s)^3 \approx \ce/\sqrt{s}$ aligned along the backbone of a 
reference chain. 
The large dashed circle represents the classical correlation hole of the total chain 
($s \approx N$) \cite{DegennesBook}.
This is the input of recent approaches to model polymer chains as soft spheres 
\cite{Likos01,Guenza04}.
We argue that incompressibility on {\em all} scales
and chain connectivity leads to a short distance repulsion of the 
segmental correlation holes, which increases with decreasing $s$. 
}
\end{figure}

\paragraph*{The scale free regime.}
Following Edwards \cite{E65,E66,DoiEdwardsBook}, the Gaussian 
(or ``Random Phase" \cite{DegennesBook}) approximation of the 
pair interaction potential in real space is
\begin{equation}
\veff(r)= v  \ \left( \delta(r)- \frac{\exp(-r/\xi)}{4\pi r\xi^2} \right)
\label{eq_v1r}
\end{equation}
where $v$ is a parameter which tunes the monomer interaction.
(It is commonly associated with the bare excluded volume of the monomers
\cite{DoiEdwardsBook}, but should more correctly be identified with the
bulk modulus effectively measured for the system. See the discussion of
Eq.~(15) of Ref.~(\cite{ANS05a}).) 
The effective potential consists of a strongly repulsive part $v \delta(r)$ of very short range, 
and a weak attractive part of range $\xi$ where the correlation length of the density fluctuations
is given by $\xi^2 = b^2 g/12$ with $g = 1/\rho v$. 
In Fourier space Eq.~(\ref{eq_v1r}) is equivalent to 
\begin{equation}
\veff(q)= v \ \frac{q^2}{q^2+\xi^{-2}}
\label{eq_v1q}
\end{equation}
with $q$ being the wave vector.
This is sufficient for calculating the scale free regime corresponding to
asymptotically long chains where chain end effects may be ignored. 
The different graphs one has to compute are indicated in Fig.~\ref{fig_diagram}. 
For $\Acal = {\bf r}^2$ (with $1 \ll n < m \ll N$) this yields, e.g.,
\begin{eqnarray}
\left< {\bf r}^2 \right> 
& = & \nonumber
b^2 \left( 1 + \frac{12}{\pi} \frac{v \xi}{b^4} \right) (m-n) - 
\frac{\sqrt{24/\pi^3}}{\rho b} \sqrt{m-n} \\
& = &
\be^2 s \left( 1 - \frac{\ce}{\sqrt{s}} \left( \frac{\be}{b} \right) \right).
\label{eq_R2pert}
\end{eqnarray}
In the second line we have used the definition of the swelling coefficient $\ce$ 
indicated in Eq.~(\ref{eq_cpce}) and have set
\begin{equation}
\be^2 \equiv b^2 \left(1 + p \frac{12}{\pi} \frac{v \xi}{b^4} \right)
= b^2 \left(1 + p \frac{\sqrt{12}}{\pi} \Gi \right)
\label{eq_bepert}
\end{equation}
with $\Gi \equiv \sqrt{v\rho} / b^3 \rho$ and $p=1$. 
(The prefactor $p$ has been added for convenience.)
The coefficient $\be$ of the leading Gaussian term in Eq.~(\ref{eq_R2pert})
--- entirely due to the graph $I_i$ describing the interactions of monomers 
inside the segment between $n$ and $m$ --- 
has been predicted long ago by Edwards \cite{DoiEdwardsBook}. It describes 
how the effective bond length is increased from $b$ to $\be$ under the 
influence of a {\em small} excluded volume interaction. 
The second term in Eq.~(\ref{eq_R2pert}) entails the $1/\sqrt{s}$-swelling
which is investigated numerically in this paper. It does only depend on
$b$ and $\rho$ but, more importantly, not on $v$ ---
in agreement with the scaling of $\Ustar(s)$ discussed in Sec.~\ref{sub_theoscal}. 
The relative weights contributing to this term are indicated in Fig.~\ref{fig_diagram}
in units of $-\sqrt{6/\pi^3} v\xi^2/b^3 \sqrt{s}$.
The diagrams $I_-$ and $I_+$ are obviously identical in the scale free limit.
Note that the interactions described by the strongest graph $I_i$ align the bonds
${\bf l}_n$ and ${\bf l}_m$ while the others tend to reduce the effect.
For higher moments of the segment size distribution $G(r,s)$ it is convenient to 
calculate first the deviation of the Fourier-Laplace transformation of $\delta G(r,s)$
and to obtain the moments from the coefficients of the expansion of this 
``generating function" in terms of the squared wave vector $q^2$. 
As explained in detail in the Appendix~\ref{app_pert} this yields more generally
\begin{equation}
\left< {\bf r}^{2p} \right> = 
\frac{(2p+1)!}{6^pp!} (\be^2 s)^p 
\left( 1 - \frac{3 (2^p p! p)^2}{2 (2p+1)!} \frac{\ce}{\sqrt{s}} 
\ \left( \frac{b}{\be} \right)^{2p-3}  \right)
\label{eq_Rppert}
\end{equation}
where we have used Eq.~(\ref{eq_bepert}) with general $p$. Obviously, Eq.~(\ref{eq_Rppert}) 
is consistent with our previous finding Eq.~(\ref{eq_R2pert}) for $p=1$. 
The corresponding segmental size distribution is
\begin{eqnarray}\nonumber
G(r,s) & = & 
\left( \frac{3}{2\pi \be^2s}\right)^{3/2} \exp\left(- \frac{3}{2} \frac{r^2}{\be^2s} \right)\\
& + & \left( \frac{3}{2\pi b^2 s} \right)^{3/2} \exp\left(- \frac{3}{2} \frac{r^2}{b^2s} \right)
\frac{\ce}{\sqrt{s}} \left(\frac{\be}{b} \right)^3 f(n)
\label{eq_Grspert}
\end{eqnarray}
with $n=r/b\sqrt{s}$ and $f(n)$ being the same function as indicated in Eq.~(\ref{eq_dGrsbe}).
The leading Gaussian terms in Eqs.~(\ref{eq_Rppert}) and (\ref{eq_Grspert}) depend on the
effective bond length $\be$, the second only on the Kuhn length $b$ of the reference chain.
%
When comparing these result with Eqs.~(\ref{eq_dGrsbe}) and (\ref{eq_Rs_key}) 
proposed in the Introduction, one sees that both equations are essentially identical --- 
taken apart, however, that they depend on $b$ and $\be$.
Note the conspicuous factor $(b/\be)^{2p-3}$ in Eq.~(\ref{eq_Rppert}) 
which would strongly reduce the empirical swelling coefficients 
$\cp = \ce (b/\be)^{2p-3}$ for large $p$ if $b$ and $\be$ were different.

\paragraph*{Interpretation of first-loop results in different contexts.}
\label{subsub_brenorm}
The above perturbation results may be used directly to describe the effect 
of a {\em weak} excluded volume $v$ on a reference system of perfectly ideal 
polymer melts with Kuhn segment length $b$ where {\em all} interactions have been switched off
($v = 0$). 
It is expected to give a good estimation for the effective bond length $\be$ 
only for a small Ginzburg parameter: $\Gi \ll 1$. For the dense melts we want to describe 
this does not hold (Sec.~\ref{sec_simu}) and one cannot hope to find a good 
quantitative agreement with Eq.~(\ref{eq_bepert}). 
Note also that large wave vectors contribute strongly
to the leading Gaussian term. The effective bond length $\be$ is, hence, 
strongly influenced by local and non-universal effects and is very difficult 
to predict in general.

Our much more modest goal is to predict the coefficient of the $1/\sqrt{s}$-perturbation 
and to express it in terms of a suitable variational reference Hamiltonian characterized
by a conveniently chosen Kuhn segment $b$ and the {\em measured} effective bond length $\be$
(instead of Eq.~(\ref{eq_bepert})). 
Following Muthukumar and Edwards~\cite{ME82}, we argue that for dense melts 
$b$ should be renormalized to $\be$ to take into account higher order graphs.
No strict mathematical proof can be given at present that the infinite number 
of possible graphs must add up in this manner. Our hypothesis relies on three
observations:
\begin{itemize}
\item
The general scaling argument discussed in Sec.~\ref{sub_theoscal} states
that we have only {\em one} relevant length scale in this problem, 
the typical segment size $R(s) \approx \be \sqrt{s}$ itself.
The incompressibility constraint cannot generate an additional scale.
It is this size $R(s)$ which sets the strength of the effective interaction
which then in turn feeds back to the deviations of $R(s)$ from Gaussianity.
Having a bond length $b$ in addition to the effective bond length $\be$ associated 
with $R(s)$ would imply incorrectly a {\em second} length scale $b \sqrt{s}$ 
varying independently with the bulk modulus $v$.
(We will check explicitly below in Fig.~\ref{fig_dGrsRes} that there is only one length scale.)
This implies $b/\be = \mbox{const} \ v^0$.
\item
Thus, since by construction $b/\be=1$ for $v\rightarrow 0$, it follows
that both lengths should be equal for all $v$.
%
%
\item
We know from Eq.~(\ref{eq_Rppert}) that the empirical coefficients $\cp = \ce (b/\be)^{2p-3}$ 
should depend strongly on the moment considered if the ratio $b/\be$ is not close to unity. 
It will be shown below (Fig.~\ref{fig_Kps}) that $\cp/\ce \approx 1$ for all $p$.
This implies $b\approx \be$.
\end{itemize}
%
%

\paragraph*{Finite chain size effects.}
To describe properly finite chain size corrections Eq.~(\ref{eq_v1q}) must be 
replaced by the general linear response formula
\begin{equation}
\frac{1}{\veff(q)} = \frac{1}{v}  + \rho F(q)
\label{eq_vFq}
\end{equation}
with $F(q) = N f_D(x) $ being the form factor of the Gaussian reference chain 
given by Debye's function $f_D(x) = 2 (e^{-x} -1 +x)/x^2$ with $x = (q b)^2 N /6$
\cite{DoiEdwardsBook}. 
This approximation allows in principle to compute, for instance, the (mean-squared)
total chain end-to-end distance, $\Acal = ({\bf r}_N - {\bf r}_1 )^2$. 
One verifies readily (see \cite{DoiEdwardsBook}, Eq.~(5.III.9)) that the effect of 
the perturbation may be expressed as
\begin{equation}
\la \Acal \ra_0 \la U \ra_0 - \la \Acal U \ra_0
=
\int \frac{d^3\bf{q}}{(2\pi)^3} \
\veff(q) \ \frac{q^2 b^4}{9} \int_0^N ds \ s^2(N-s)\exp\left(-\frac{q^2b^2s}{6}\right).
\label{eq_Rendcorrect1}
\end{equation}
We take now first the integral over $s$. In the remaining integral over $q$
small $q$ wave vectors contribute to the $\sqrt{N}$-swelling while large 
$q$ renormalize the effective bond length of the dominant Gaussian 
behaviour linear in $N$ (as discussed above). Since we wish to determine 
the non-Gaussian corrections, we may focus on small wave vectors $q \ll 1/\xi$.
Since in this limit $1/v = \rho g \ll \rho F(q)$, 
one can neglect in Eq.~(\ref{eq_vFq}) the $1/v$ contribution 
to the inverse effective interaction potential.
We thus continue the calculation using the much simpler $\veff(N,x) = 1/(N\rho f_D (x))$.
This allows us to express the swelling as 
\begin{equation}
1 - \frac{\left< ({\bf r}_N - {\bf r}_1)^2 \right> }{\be^2 N} =
\frac{\ce}{\sqrt{N}} \ I(x_u).
\label{eq_Rendcorrect2}
\end{equation}
To simplify the notation we have set here finally $b=\be$ in agreement with the 
hypothesis discussed above. The numerical integral 
$I(x_u) = \int_0^{x_u} dx \ldots$ 
over $x$ is slowly convergent at infinity. As a consequence the estimate 
$I(\infty)=1.59$ may be too large for moderate 
chain lengths. In practice, convergence is not achieved for values 
$x_u(N) \approx (b/\xi)^2 N$ corresponding to the screening length $\xi$.

We remark finally that numerical integration can be avoided for various properties 
if the Pad\'e approximation of the form factor, $F(q) = N /(1+(q b)^2 N/12)$, 
is used. This allows analytical calculations by means of the simplified
effective interaction potential 
\begin{equation}
\veff(q) = \frac{q^2b^5}{12\rho b^3}+\frac{b^3}{N\rho b^3}.
\label{eq_vPade}
\end{equation}
This has been used for instance for the calculation of finite chain size effects
for the bond-bond correlation function discussed in 
Sec.~\ref{subsub_Ps_finiteN} below
\footnote{It is interesting to compare the numerical value 
$I(\infty) \approx 1.59$ obtained for the  {\em r.h.s} of Eq.~(\ref{eq_Rendcorrect2}) 
with the coefficients one would obtain by computing Eq.~(\ref{eq_Rendcorrect1}) either 
with the effective potential $\veff(q)$ for infinite chains given by Eq.~(\ref{eq_v1q})
or with the Pad\'e approximation, Eq.~(\ref{eq_vPade}). Within these approximations of
the full linear response formula, Eq.~(\ref{eq_vFq}), the coefficients can be obtained
directly without numerical integration yielding overall similar values. In the first case
we obtain $15/8 \approx 1.87$ and in the second $11/8 \approx 1.37$. 
While the first value is clearly not compatible with the measured end-to-end distances, 
the second yields a reasonable fit, especially for small $N < 1000$, when the
data is plotted as in Fig.~\ref{fig_K1s}. Ultimately, for very long chains the 
correct coefficient should be $1.59$ as is indicated by the dash-dotted line
in Fig.~\ref{fig_Rslog}.
}.

\section{Computational models and technical details\label{sec_simu}}

\subsection{Bond fluctuation model\label{sub_BFM}}

\paragraph*{A widely-used lattice Monte Carlo scheme for coarse-grained polymers.}
The body of our numerical data comes from the three dimensional bond fluctuation model (BFM) ---
a lattice Monte Carlo (MC) algorithm where each monomer occupies  eight sites of a unit cell of 
a simple cubic lattice \cite{BFM,Deutsch,Paul91a}. Our version of the BFM with $108$ bond vectors 
corresponds to flexible athermal chain configurations \cite{BWM04}. All length scales are given 
in units of the lattice constant and time in units of Monte Carlo Steps (MCS).
We use cubic periodic simulation boxes of linear size $L=256$ containing 
$\nmon =\rho L^3=2^{20} \approx 10^6$ monomers. This monomer number corresponds 
to a monomer number density $\rho=0.5/8$ where half of the lattice sites are occupied 
(volume fraction $0.5$). 
The large system sizes used allow us to suppress finite box-size effects for 
systems with large chains. 
Using a mix of local, slithering snake \cite{Kron65,Wall75,MWBBL03}, and double-bridging 
\cite{Theodorou02,dePablo03,Auhl03,BWM04} MC moves we were able to equilibrate dense systems 
with chain lengths up to $N=8192$. 
%
%

\begin{figure}[tb]
\includegraphics*[width=0.9\textwidth]{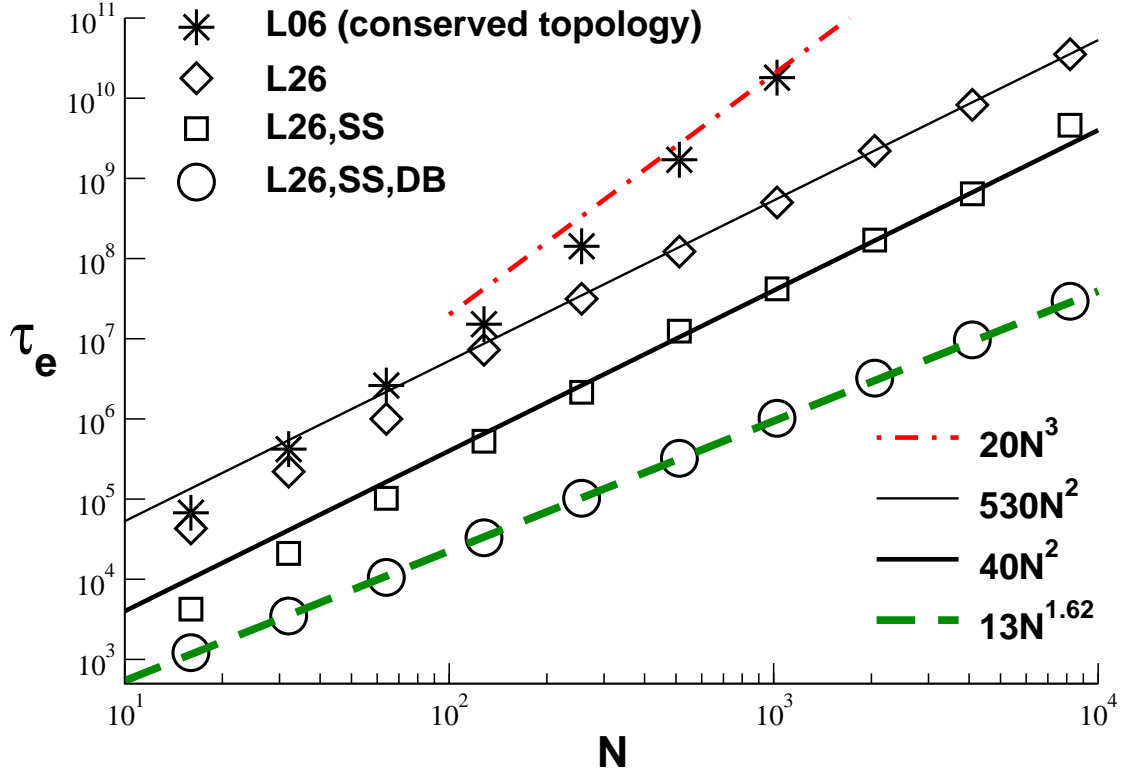}
\caption{(Color online)
Diffusion time $\taue$ over the (root-mean-squared) chain 
end-to-end distance $\Rend(N-1)$ as a function of chain length $N$ for 
different versions of the Bond Fluctuation Model (BFM).
All data indicated are for the high number density ($\rho=0.5/8$) corresponding 
to a polymer melt with half the lattice sites being occupied.  
We have obtained $\taue = \Rend^2(N-1)/6 \Ds$ from the self-diffusion coefficient $\Ds$ 
measured from the free diffusion limit of the mean-squared displacement of all 
monomers $\left< \delta r(t)^2 \right> = 6 \Ds t$. 
Data from the classical BFM with topology conserving local Monte Carlo (MC) moves in
6 spatial directions (L06) \cite{Paul91a} are represented as stars. All other data
sets use topology violating local MC moves in 26 lattice directions (L26).
If only local moves are used, L26-dynamics is even at relatively short times perfectly Rouse like
which allows the accurate determination of $\Ds$ although the monomers 
possibly have not yet moved over $\Rend(N-1)$ for the largest chain lengths considered. 
Additional slithering snake (SS) moves increase the efficiency of the algorithm
by approximately an order of magnitude (squares,bold line).
The power law exponent is changed from $2$ to an empirical $1.62$
(dashed line) if in addition we perform double-bridging (DB) moves.
\label{fig_taue}
}
\end{figure}

\paragraph*{Equilibration and sampling of high-molecular BFM melts.}
%
Standard BFM implementations \cite{Paul91a,MWB00,KBMB01} use local MC jumps to the 6 closest 
lattice sites to prevent the crossing of chains and conserve therefore the chain topology.
These ``L06" moves lead to very large relaxation times, scaling at least as $\taue \sim N^3$,
as may be seen from Fig.~\ref{fig_taue} (stars). The relaxation time $\taue = \Rend^2/6\Ds$ 
indicated in this figure has been estimated from the self-diffusion coefficient $\Ds$ obtained 
from the mean-square displacements of all monomers in the free diffusion limit. (For the largest
chain indicated for L06 dynamics only a lower bound for $\taue$ is given.) 
Instead of this more realistic but very slow dynamical scheme we make jump attempts to the 26 sites 
of the cube surrounding the current monomer position (called ``L26" moves).
This allows the chains to cross each other which dramatically speeds up the dynamics, 
especially for long chains ($N > 512$). 
If only local moves are considered, the dynamics is perfectly consistent with the Rouse model 
\cite{DoiEdwardsBook}. As shown in Fig.~\ref{fig_taue}, we find $\taue \approx 530 N^2$ for 
L26 dynamics. This is, however, still prohibitive by large for sampling configurations 
with the longest chain length $N$ we aim to characterize 
\footnote{These topology non-conserving moves yield configurations which are not accessible 
with the classical scheme with jumps in 6 directions only. Concerning the {\em static}
properties we are interested in this paper both system classes are practically equivalent.
This has been confirmed by comparing various static properties and by counting the
number of monomers which become ``blocked" (in absolute space or with respect to an 
initial group of neighbor monomers) once one returns to the original local scheme.
Typically we find about 10 blocked monomers for a system of $2^{20}$ monomers.
The relative difference of microstates is therefore tiny and irrelevant for
static properties. Care is needed, however, if the equilibrated configurations
are used to investigate the {\em dynamics} of the topology conserving BFM version. 
The same caveats arise for the slithering snake and double-bridging moves.
}.

\paragraph*{Slithering snake moves.}
In addition to the local moves {\em one} slithering snake move per chain is attempted on
average per MCS corresponding to the displacement of $N$ monomer along the chain backbone. 
Note that in our units {\em two} spatial displacement attempts per MCS are performed
on average per monomer, one for a local move and one for a snake move.
(In practice, it is computationally more efficient for large $N$ to take off a 
monomer at one chain end and to paste it at the other leaving all other monomers
unaltered. Before dynamical measurements are performed the original order of
beads must then be restored.) 
Interestingly, a significantly larger slithering snake attempt frequency would
not be useful since the relaxation time of slithering snakes without or only few 
local moves increases exponentially with mass \cite{Deutsch85,MWBBL03} due to the 
correlated motion of snakes \cite{ANS97}. In order to obtain an efficient free 
snake diffusion (with a chain length independent curvilinear diffusion coefficient 
$\Dc(N) \sim N^0$ and $\taue \approx N^2/\Dc(N) \sim N^2$ \cite{Wall75,MWBBL03}) 
it is important to relax density fluctuations rapidly by local dynamical pathways. 
As shown in Fig.~\ref{fig_taue} (squares), we find a much reduced relaxation time 
$\taue \approx 40 N^2$ which is, however, still unconveniently large for our longest chains.
Note that most of the CPU time is still used by local moves.
The computational 
load per MCS remains therefore essentially chain length independent.

\paragraph*{Advantages and pitfalls of double-bridging moves.}
Double-bridging (DB) moves are very useful for high densities and help us to extend the
accessible molecular masses close to $10^4$. As for slithering snake moves we use
all $108$ bond vectors to switch chain segments between two different chains. 
Only chain segments of equal length are swapped to conserve monodispersity. 
Topolocial constraints are again systematically and deliberately violated. 
Since more than one swap partner is possible for a selected first monomer, 
delicate detailed balance questions arise. This is particularly important for
short chains and is discussed in detail in Ref.~\cite{BWM04}. Technically,
the simplest solution to this problem is to refuse all moves with more than 
one swap partner (to be checked both for forward and back move). 
The configurations are screened with a frequency $f_\text{DB}$ for possible DB moves
where we scan in random order over the monomers. The frequency should not
be too large to avoid (more or less) immediate back swaps and monomers should
move at least out of the local monomer cage and over a couple of lattice sites. 
We use $f_\text{DB}=0.1$ between DB updates for the configurations reported here.
(The influence of $f_\text{DB}$ on the performance has not been explored systematically, 
but preliminary results suggest a slightly smaller DB frequency for future studies.)
The diffusion times over the end-to-end distance for this case are indicated 
in Tab.~\ref{tab_BFMmelt}. As shown in Fig.~\ref{fig_taue}, we find
empirically $\taue(N) \approx 13 N^{1.62}$.
For $N=8192$ this corresponds to $3 \cdot 10^7$ MCS.
This allows us even for the largest chain lengths to observe monomer
diffusion over several $\Rend$ within the $10^8$ MCS which are 
feasible on our XEON-PC processor cluster.

The efficiency of DB moves is commonly characterized in terms of the relaxation 
time $\tau_\text{ee}$ of the end-to-end vector correlation function \cite{Theodorou02,dePablo03}. 
For normal chain dynamics this would indeed characterize the longest relaxation time 
of the system, i.e. $\taue \approx \tau_\text{ee}$. 
For the double-bridging this is, however, not sufficient since density fluctuations do 
{\em not} couple to the bridging moves and can not be relaxed. We find therefore that 
configurations 
equilibrate on time scales given by $\taue$ rather than by $\tau_\text{ee} \ll \taue$.
This may be verified, for instance, from the time needed for the distribution
$\Rend(s)$ (and especially its spatial components) to equilibrate. 
The criterion given in the literature \cite{Theodorou02} is clearly 
not satisfactory and may lead to insufficiently equilibrated configurations.
In summary, equilibration with DB moves still requires monomer diffusion over the 
typical chain size, however at a much reduced price.

%

\paragraph*{Some properties of our configurations.}
%
The Tables~\ref{tab_BFMmelt} and \ref{tab_BFMBSMmelt} summarize some system 
properties obtained for our reference density $\rho=0.5/8$.  
Averages are performed over all chains and $1000$ configurations.
These configurations may be considered to be independent for $N < 4096$.
Only a few independent configurations exist for the largest chain length $N=8192$
which has to be considered with some care. Taking apart this system, chains are always
much smaller than the box size.
For asymptotically long chains, we obtain an average bond length 
$\la |{\bf l}| \ra \approx 2.604$, 
a root-mean-squared bond length $l \equiv \la {\bf l}^2 \ra^{1/2} \approx 2.635$
and an effective bond length $\be \approx 3.244$
--- as we will determine below in Sec.~\ref{sub_Rs}.
This corresponds to a ratio $C_{\infty} \equiv \be^2/l^2 \approx 1.52$
and, hence, to a persistence length $\lp = l (C_{\infty}+1)/2 \approx 3.32$ \cite{BWM04}.
Especially, we find from the zero wave vector limit of the total structure factor $S(q)$ 
a low (dimensionless) compressibility 
$g = S(q\rightarrow 0)/\rho \approx 0.246$ which compares well with real experimental melts.
From the measured bulk compression modulus $v \equiv 1/g(\rho)\rho$ and the 
effective bond length $\be$ one may estimate a Ginzburg parameter 
$\Gi = \sqrt{v\rho} / \be^3 \rho \approx 0.96$.
Following Ref.~\cite{ANS05a} the interaction parameter $v$ is supposed here to
be given by the full inverse compressibility and not just by the second virial 
coefficient. 

\subsection{Bead spring model\label{sub_BSM}}

\paragraph*{Hamiltonian.}
Additionally, molecular dynamics simulations of a bead-spring model (BSM) \cite{MMP01} 
were performed to dispel concerns that our results are influenced by the underlying 
lattice structure of the BFM. The model is derived from a coarse-grained model for
polyvinylalcohol which has been employed to study polymer crystallization \cite{MMP02}.
It is characterized by two potentials: a non-bonded potential of Lennard-Jones (LJ) type
and a harmonic bond potential. 
While the often employed Kremer-Grest model \cite{KG90} uses a $12-6$ LJ potential to 
describe the non-bonded interactions $U_\text{nb}(r)$, our non-bonded potential has a 
softer repulsive part. It is given by
\begin{equation}
U_\text{nb}(r) = 1.511 \left[\left(\frac{\sigma_0}{r}\right)^9-\left(\frac{\sigma_0}{r}\right)^6\right],
\label{eq_BSMnonbonded}
\end{equation}  
which is truncated and shifted at the minimum at $r_\text{min} \approx 1.15$.
Note that all length scales are given in units of $\sigma_0$ and we use LJ units 
\cite{AllenTildesleyBook}
for all BSM data (mass $m=1$, Boltzmann constant $\kB=1$).
The parameters of the bond potential, $U_\text{b}(r) = 1120 (r-l_\text{b})^2$, 
are adjusted so that the average
bond length $l(\rho=0.84)\approx l_\text{b} = 0.97$ is approximately 
the same as in the standard Kremer-Grest model \cite{KG90}. 
The average bond length and the root-mean-squared bond length 
are almost identical for the BSM due to the very stiff bond potential.
Since $r_\text{min}/l \approx 1.16$ bonded monomers penetrate each other significantly.

\paragraph*{Equilibration and sampling.}
We perform standard molecular dynamics simulations in the canonical ensemble with a
Langevin thermostat (friction constant $\Gamma=0.5$) at temperature $T=1$.
The equations of motion are integrated by the velocity-Verlet algorithm \cite{AllenTildesleyBook}.
To improve the statistics for large chain length, we have implemented 
additional double-bridging moves. Since only few of these MC moves are accepted 
per unit time, this does affect neither the stability nor the accuracy of the 
molecular dynamics sweeps.

\paragraph*{Some properties obtained.}
For clarity, we show only data for chain length $N=1024$ and number density 
$\rho = 0.84$, the typical melt density of the Kremer-Grest model \cite{KG90}. 
For the reported data we use periodic simulation boxes of linear size $L \approx 62$ 
containing $\nmon = 196608$ monomers, but we have also sampled different boxes sizes
(up to $L=77.5$) to check for finite box-size effects.
For the reference density a dimensionless compressibility $g \approx 0.08$ is found
which is about three times smaller than for our BFM melt.
For the effective bond length we obtain $\be \approx 1.34$, i.e. 
BSM chains ($C_{\infty} \approx 1.91$, $\lp \approx 1.41$) are slightly stiffer
than the corresponding BFM polymers.
%
%
Fortunately, the product $\rho \be^3 \approx 2$ is roughly similar in both models 
and one expects from Eq.~(\ref{eq_cpce}) a similar swelling for large $s$.
Note finally that the Ginzburg parameter $\Gi \approx 1.8$ is much 
larger than for the BFM systems. As we have emphasized in Sec.~\ref{sec_theo},
this should, however, not influence the validity of the perturbation prediction of 
the expected $1/\sqrt{s}$-swelling of the chains when expressed in terms of the 
{\em measured} effective bond length.

\section{Numerical results}
\label{sec_resu}

As illustrated in Fig.~\ref{fig_diagram}, a chain segment of curvilinear 
length $s > 0$ is identified by two monomers $n$ and $m=n+s$ on the same chain.
We compute here various moments of chain segment properties where we ensemble-average 
over all chains and all start points $n$.
The statistical accuracy must therefore always decrease for large $s$.
We concentrate first on the second moment ($p=1$) of the segmental size distribution.
Higher moments and the segmental size distribution are discussed in Sec.~\ref{sub_Grs}.

\subsection{The swelling of chain segments}
\label{sub_Rs}

\begin{figure}[tb]
\includegraphics*[width=0.9\textwidth]{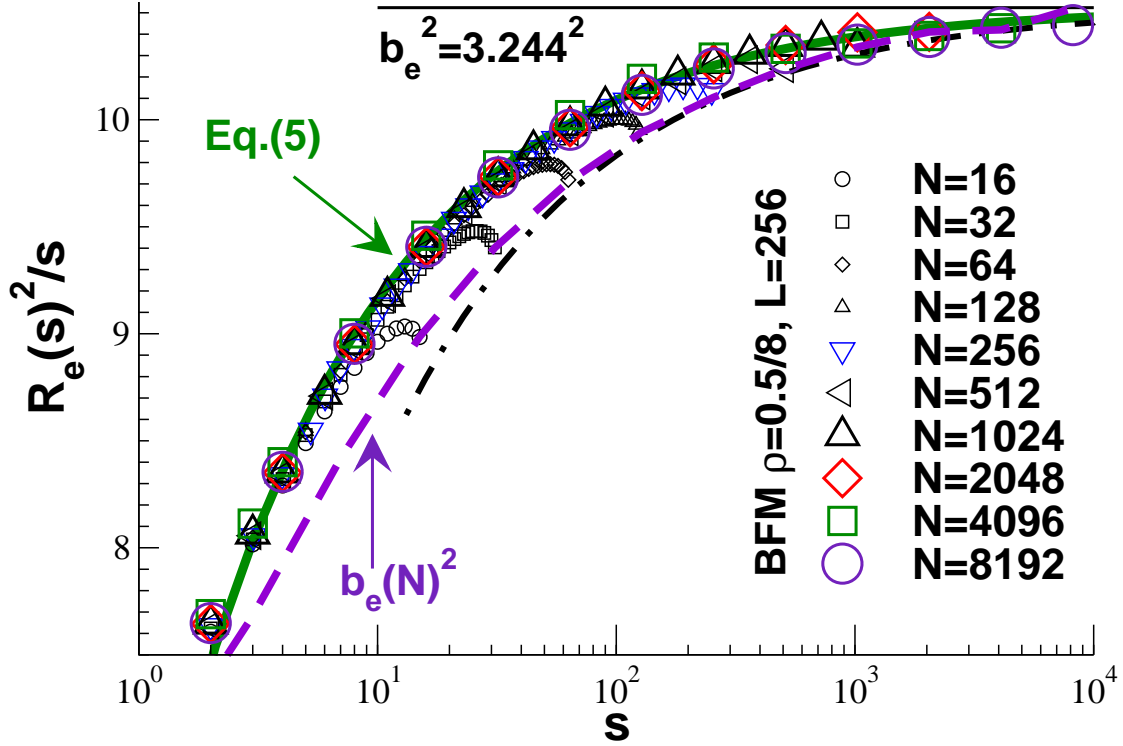}
\caption{(Color online)
\label{fig_Rslog}
Mean-squared segment size $\Rend(s)^2/s$ {\em vs.} curvilinear distance $s$.
We present BFM data for different chain length $N$ at number density $\rho =0.5/8$.
The averages are taken over all possible monomer pairs $(n,m=n+s)$.
The statistics deteriorates, hence, for large $s$.
Log-linear coordinates are used to emphasize the power law swelling over several 
orders of magnitude of $s$. The data approach the asymptotic limit (horizontal line) 
from below, i.e. the chains are {\em swollen}. This behavior is well fitted by 
Eq.~(\ref{eq_Rs_key}) for $1 \ll s \ll N$ (bold line). 
Non-monotonic behavior is found for $s \rightarrow N$, especially for small $N$. 
The dashed line indicates the measured total chain end-to-end distances, 
$\be(N)^2 \equiv \Rend(N-1)^2/(N-1)$ from Tab.~\ref{tab_BFMmelt},
showing even more pronounced deviations from the asymptotic limit.
The dash-dotted line compares this data with Eq.~(\ref{eq_ReN}).
}
\end{figure}

\begin{figure}[tb]
\includegraphics*[width=0.9\textwidth]{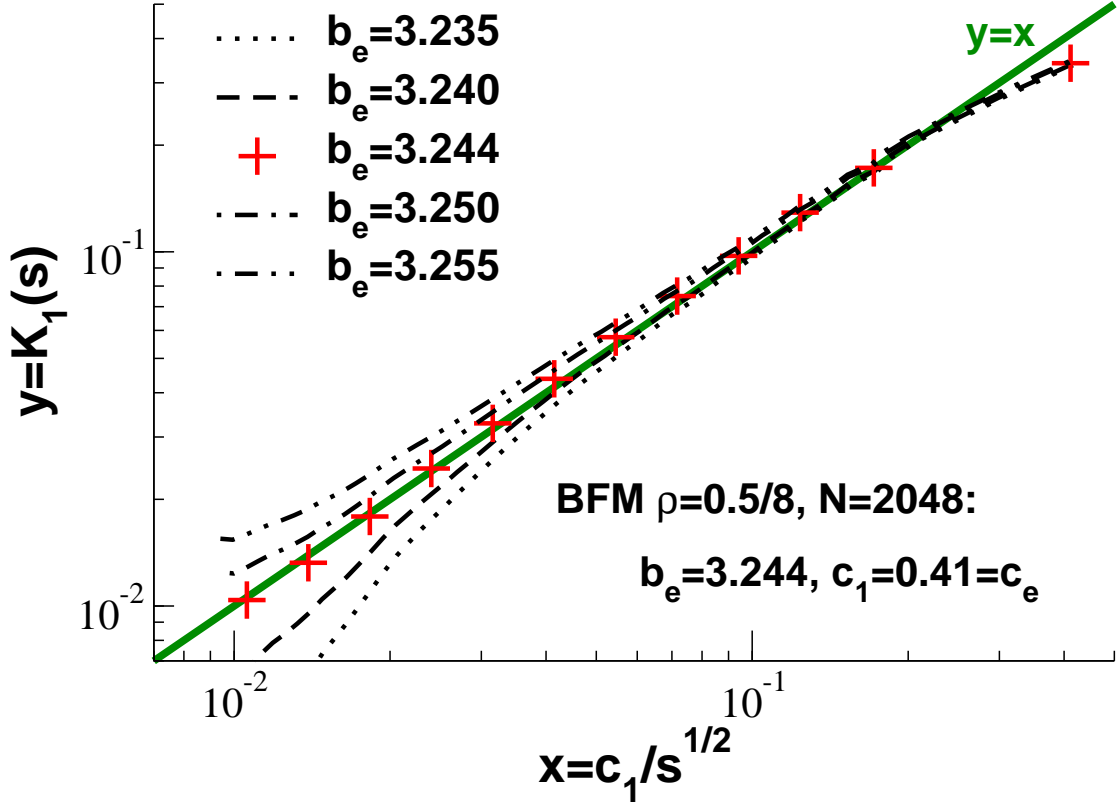}
\caption{(Color online)
\label{fig_K1s}
Replot of the mean-squared segment size as $y = K_1(s) = 1 - \Rend(s)^2 /\be^2s$ 
{\em vs.} $x = c_1/\sqrt{s}$, as suggested by Eq.~(\ref{eq_Rs_key}),
for different trial effective bond lengths $\be$ as indicated.
Only BFM chains of length $N=2048$ are considered for clarity. 
This procedure is very sensitive to the value chosen and allows for a precise determination.
It assumes, however, that higher order terms in the expansion of $K_1(s)$ may be neglected. 
The value $\be$ is confirmed from a similar test for higher moments (Fig.~\ref{fig_Kps}).
}
\end{figure}

\begin{figure}[tb]
\includegraphics*[width=0.9\textwidth]{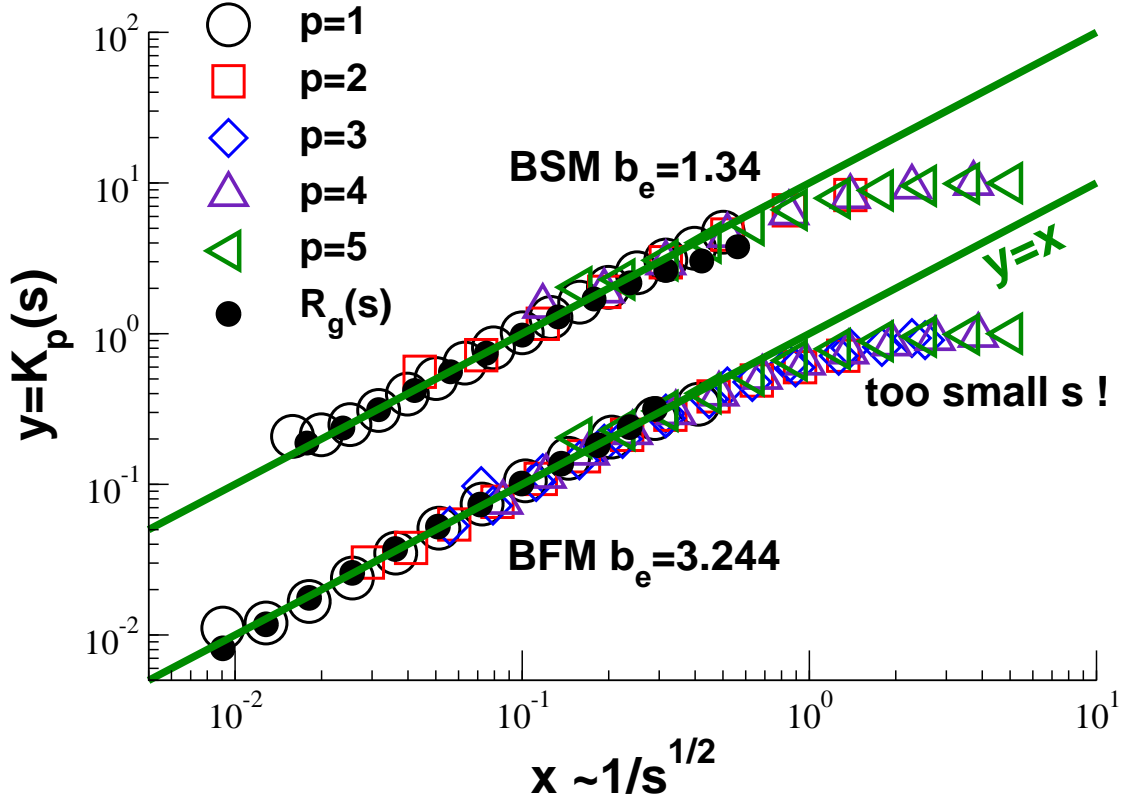}
\caption{(Color online)
\label{fig_Kps}
Critial test of Eq.~(\ref{eq_Rs_key}) where the rescaled moments $y=\Kp(s)$ 
of the segment size distribution (defined in Eq.~(\ref{eq_Kpdef})) are
plotted {\em vs.} $x = \frac{3 (2^p p! p)^2}{2 (2p+1)!} \ \frac{\cp}{\sqrt{s}}$.
We consider the first five even moments ($p=1,\ldots,5$) for the BFM with $N=2048$ 
and the BSM with $N=1024$. Also indicated is the rescaled radius of gyration, 
$y = 5/8 \ ( 1-6\Rgyr^2(s)/\be^2 (s+1))$, as a function of $x = c_1/\sqrt{s+1}$
(filled circles). The BSM data has been shifted upwards for clarity.
Without this shift a perfect data collapse is found for both models and all moments.
Keeping the same effective bond length $\be$ for all moments of each model
we fit for the swelling coefficients $\cp$ by rescaling the horizontal axis.
We find $\be \approx 3.244$ for the BFM and $1.34$ for the BSM. If $\be$
is chosen correctly, all data sets extrapolate linearly to zero for large $s$
($x \rightarrow 0$). The swelling coefficients found are close the theoretical
prediction $\ce$, as indicated in Tab.~\ref{tab_BFMBSMmelt}.
The plot demonstrates that the non-Gaussian deviations scale as the segmental
correlation hole, $\Ustar(s) \sim \ce/\sqrt{s}$ and this for all moments as long as $x \ll 1$.
The saturation at large $x$ is due to the finite extensibility of short chain segments.
Since this effect becomes more marked for larger moments, the fit of $\be$ is
best performed for $p=1$.
}
\end{figure}

\paragraph*{Scale free regime for $1 \ll s \ll N$.}
The mean-squared segment size $\Rend^2(s) = \la {\bf r}^2\ra$ is presented in 
the Figs.~\ref{fig_Rslog}, \ref{fig_K1s} and \ref{fig_Kps}. The first plot shows 
clearly that chain segments are swollen, i.e. $\Rend^2(s)/s$ increases systematically
and this up to very large curvilinear distances $s$. Only BFM data are shown for clarity. 
A similar plots exists for the BSM data. 
In agreement with Eq.~(\ref{eq_Rs_key}) for $p=1$, the asymptotic Gaussian behavior 
(dashed line) is approached from below and the deviation decays as 
$\Ustar(s) \propto 1/\sqrt{s}$ (bold line). 
The bold line indicated corresponds to $\be = 3.244$ 
and $c_1 \approx \ce \approx 0.41$ 
which fits nicely the data over several decades in $s$ ---
provided that chain end effects can be neglected ($s \ll N$). 
Note that a systematic {\em underestimation} of the true effective bond length 
would be obtained by taking simply the largest $\Rend^2(s)/s \approx 3.23^2$ value 
available, say, for monodisperse chains of length $N=2048$.
\paragraph*{Finite chain-size effects.}
Interestingly, $\Rend^2(s)/s$ does not approach the asymptotic limit monotonicly. 
Especially for short chains one finds a {\em non-monotonic} behavior for $s \rightarrow N$.
This means that the total chain end-to-end distance $\Rend(s=N-1)$ must show even more 
pronounced deviations from the asymptotic limit. This is confirmed by the dashed line
representing the $\be^2(N) \equiv \Rend^2(N-1)/(N-1)$ data points given in Tab.~\ref{tab_BFMmelt}.
We emphasize that the non-monotonicity of $\Rend^2(s)/s$  becomes weaker with increasing $N$ 
and that, as one expects, the inner distances, as well as the total chain size, 
are characterized by the {\em same} effective bond length $\be$ for large $s$ or $N$.
The non-monotonic behavior may be qualitatively understood by the reduced 
self-interactions at chain ends which lessens the swelling on these scales. 
These finite-$N$ corrections have been calculated analytically using the full 
Debye function for the effective interaction potential $\veff(q)$, Eq.~(\ref{eq_vFq}). 
The prediction for the total chain end-to-end vector given in Eq.~(\ref{eq_Rendcorrect2}) 
is indicated in Fig.~\ref{fig_Rslog} (dash-dotted line) where we have replaced
the weakly $N$-dependent integral $I(x_u)$ by its upper bound value for infinite chains
\begin{equation}
1 - \frac{\Rend^2(N-1)}{\be^2 \ (N-1)} = 
\frac{1.59 \ce}{\sqrt{N-1}}.
\label{eq_ReN}
\end{equation}
We have changed here the chain length $N$ in the analytical formula (obtained for
large chains where $N \approx N-1$) to the curvilinear length $N-1$. This is physically 
reasonable and allows to take better into account the behavior of small chains. 
Note that Eq.~(\ref{eq_ReN}) is similar to Eq.~(\ref{eq_Rs_key}) 
--- apart from a slightly larger prefactor explaining the 
observed stronger deviations. 
Theory compares well with the measured data for large $N$. It does less so for smaller $N$, 
as expected, where the chain length dependence of the numerical integral $I(x_u(N)) \le 1.59$ 
must become visible. This explains why the data points are {\em above} the dash-dotted line. 
Note also that additional non-universal finite-$N$ effects not accounted for by the theory
are likely for small $N$. 
In contrast to this, $\Rend(s)$ is well described by the theory 
even for rather small $s$ provided that $N$ is large and chain end effects can be neglected.
In summary, it is clear that one should use the segment size $\Rend(s)$ rather
than the total chain size to obtain in a computational study a reliable fit of 
the effective bond length $\be$.

%
%
%
%
\paragraph*{Extrapolation of the effective bond length of asymptotically long chains.}
The representation chosen in Fig.~\ref{fig_Rslog} is not the most convenient one
for an accurate determination of $\be$ and $c_1$. How precise coefficients may be
obtained according to Eq.~(\ref{eq_Rs_key}) is addressed in the Figs.~\ref{fig_K1s} 
and \ref{fig_Kps}. 
The fitting of the effective bond length $\be$ and its accuracy is illustrated in 
Fig.~\ref{fig_K1s} for BFM chains of length $N=2048$. This may be first done 
approximately in linear coordinates by plotting $\Rend^2(s)/s$ as a function of 
$1/\sqrt{s}$ (not shown). Since data for large $s$ are less visible in this 
representation, we recommend for the fine-tuning of $\be$ to switch then to 
logarithmic coordinates with a vertical axis $y=1 - \Rend^2(s)/ \be^2 s$ for 
different trial values of $\be$.
The correct value of $\be$ is found by adjusting the vertical axis $y$ such that 
the data extrapolates {\em linearly} as a function of $1/\sqrt{s}$ to zero for large $s$. 
We assume for the fine-tuning that higher order perturbation corrections may be neglected, 
i.e. we take Eq.~(\ref{eq_Rs_key}) literally. (We show below that higher order corrections 
must indeed be very small.) The plot shows that this method is very sensitive, yielding a 
best value that agrees with the theory over more than one order of magnitude {\em without} 
curvature.
As expected, it is not possible to rationalize the numerically obtained
values $\be \approx 3.244$ for the BFM and $\be \approx 1.34$ for the BSM
using Eq.~(\ref{eq_bepert}).
According to Eq.~(\ref{eq_cpce}) these fit values imply the theoretical
swelling coefficients $\ce = 0.41 $ for the BFM and $\ce = 0.44 $ for the BSM.
\paragraph*{Empirical swelling coefficients.}
As a next step the horizontal axis is rescaled such that all data sets collapse 
on the bisection line, i.e. using Eq.~(\ref{eq_Rs_key}) we fit for the empirical 
swelling coefficient $c_1$ and compare it to the predicted value $\ce$.
This rescaling of the axes allows to compare both models in Fig.~\ref{fig_Kps}. 
For clarity the BSM data have been shifted upwards.
For the BFM we find $c_1/\ce \approx 1.0$, as expected, while our BSM simulations yield a 
slightly more pronounced swelling with $c_1/\ce \approx 1.2$.

\paragraph*{Segmental radius of gyration.}
Also indicated in Fig.~\ref{fig_Kps} is the segmental radius of gyration $\Rgyr(s)$ 
(filled circles) computed as usual \cite{DoiEdwardsBook} as the variance of the positions 
of the segment monomers around their center of mass. Being the sum over all $s+1$ 
monomers, it has a much better statistics compared to $\Rend(s)$. 
The scaling used can be understood by expressing the radius of gyration 
$\Rgyr^2(s) = \frac{1}{(s+1)^2} \la \sum_{k=n}^{n+s} \sum_{l=n}^{n+s} {\bf r}_{kl}^2 \ra$
in terms of displacement vectors ${\bf r}_{kl}$ \cite{DoiEdwardsBook}. 
Using Eq.~(\ref{eq_Rs_key}) and integrating twice this yields
\begin{equation}
1 - \frac{6 \Rgyr^2(s)}{\be^2 (s+1)} = \frac{8}{5} \frac{c_1}{\sqrt{s+1}}.
\label{eq_Rgs}
\end{equation}
Plotting the {\em l.h.s.} of this relation against the {\em r.h.s.} we obtain a
perfect data collapse on the bisection line where we have used the
{\em same} parameters $\be$ and $c_1$ as for the mean-squared segment size.
This is an important cross-check which we strongly recommend. 
Different values indicate insufficient sample equilibration.

\subsection{Chain connectivity and recursion relation}

\begin{figure}[tb]
\includegraphics*[width=0.9\textwidth]{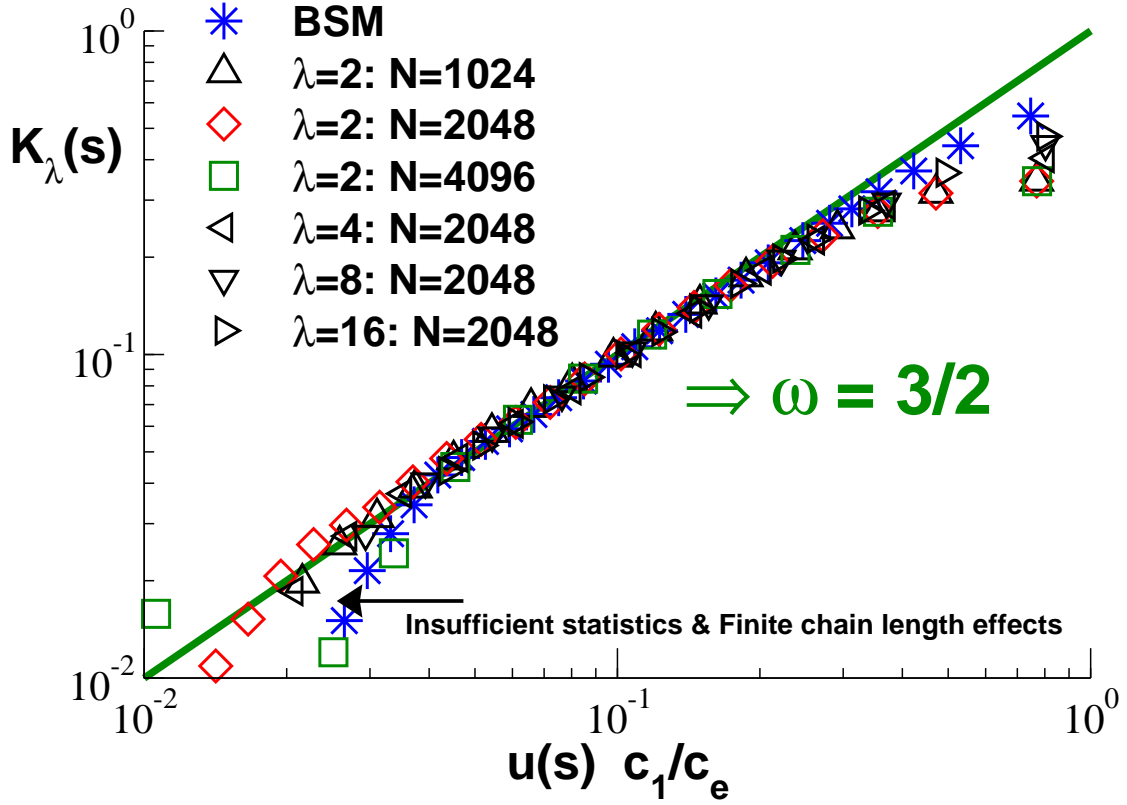}
\caption{(Color online)
Plot of $K_{\lambda}(s)$ as a function of $\Ustar(s) c_1/\ce \sim 1/\sqrt{s}$
using the {\em measured} $\Ustar(s) \equiv \sqrt{24/\pi^3} s/\rho \Rend(s)^3$.
For $\lambda=2$ (corresponding to two segments being connected) BFM and BSM data
are compared. Several $\lambda$ values are given for $N=2048$ BFM chains.
For chain segments with $1 \ll s \ll N$ all data sets collapse on the bisection line 
confirming the so-called ``recursion relation" $K_{\lambda} \approx \Ustar$ proposed 
by Semenov and Johner \cite{ANS03}. 
The statistics becomes insufficient for large $s$ (left bottom corner). 
Systematic deviations arise for $s \rightarrow N$ due to additional finite-$N$ effects.
\label{fig_Krecurs}
}
\end{figure}

As was emphasized in Sec.~\ref{sub_theoscal} the observed swelling is due to an 
entropic repulsion between chain segments induced by the incompressibility
of the melt. To stress the role of {\em chain connectivity} we repeat the general 
scaling argument given above in a form originally proposed by Semenov and Johner 
for ultrathin films \cite{ANS03}. As shown in Fig.~\ref{fig_Krecurs} we test the
relation 
\begin{equation}
K_{\lambda}(s) \equiv \frac{\Rend^2(\lambda s) - \lambda \Rend^2(s)}{(\lambda-\sqrt{\lambda}) \Rend^2(s)} \approx \Ustar(s) 
\equiv \frac{s}{\rho \Rend(s)^d} 
\label{eq_recursion}
\end{equation}
with $K_{\lambda}(s)$ being a direct measure of the non-Gaussianity
($\lambda$ being a positive number) 
comparing the size of a segment of length $\lambda s$ with the 
size of $\lambda$ segments of length $s$ joined together.
%
%
(The prefactor $1/(\lambda-\sqrt{\lambda})$ in the definition of $K_{\lambda}(s)$
has been introduced for convenience.)
Equivalently, this can be read as a measure for the swelling of a chain
where initially the interaction energy $\Ustar$ between the segments
has been switched off. $K_{\lambda}$ is a functional of $\Ustar(s)$ 
with $K_{\lambda}[\Ustar =0]=0$. The analytic expansion of the functional 
must be dominated by the linear term (as indicated by $\approx$ in the above relation)
simply because $\Ustar$ is very small.
Altogether, Eq.~(\ref{eq_recursion}) yields a {\em recursion relation} relating
$\Rend(\lambda s)$ with $\Rend(s)$ for any $\lambda$ provided $1 \ll s < \lambda s \ll N$.
It can be solved, leading (in lowest order) to Eq.~(\ref{eq_Rs_key}) with $p=1$.
This may be seen from the ansatz $\Rend^2(s) = \be^2 s (1 - \ce/ s^{\omega-1} + \ldots)$
which readily yields $\omega = 3 /2$ and $\ce \approx 1/\rho \be^3$.
%
%

Eq.~(\ref{eq_recursion}) has been validated directly in Fig.~\ref{fig_Krecurs}
for $\lambda=2$ (corresponding to two segments of length $s$ joined together) 
for the BFM and the BSM as indicated. In addition, for BFM chains of length $N=2048$ 
several values of $\lambda$ have been given. As suggested by Eq.~(\ref{eq_cpce}),
we have plotted $K_{\lambda}(s)$ as a function of $(c_1/\ce) \ \Ustar(s)$
with $\Ustar(s) \equiv \sqrt{24/\pi^3} s/\rho \Rend^3(s) \approx \ce/\sqrt{s}$.
The prefactor of $\Ustar(s)$ allows a convenient comparison with Fig.~\ref{fig_K1s}.
Note the perfect data collapse for all data sets.
More importantly, the predicted linearity is well confirmed for large segments ($1 \ll s$)
and this without any tunable parameter for the vertical axis, as was needed in the
previous Figs.~\ref{fig_K1s} and \ref{fig_Kps}.
%
%

\subsection{Intrachain bond-bond correlations \label{sub_Ps}}

\paragraph*{Expectation from Flory's hypothesis.}
\label{subsub_Ps}
An even more striking violation of Flory's ideality hypothesis may be obtained 
by computing the bond-bond correlation function, defined by the first Legendre 
polynomial 
$P(s)= \la {\bf l}_{m=n+s} \cdot {\bf l}_n \ra / l^2$
where the average is performed, as before, over all possible pairs of monomers $(n,m=n+s)$
\footnote{We have verified that the alternative definition
$\la {\bf e}_{m=n+s} \cdot {\bf e}_n \ra$ 
with ${\bf e}_n$ being the normalized bond vector yields very similar results.
This is due to the weak bond length fluctuations, specifically at high densities,
in both coarse-grained models under consideration. It is possible that other models
show slightly different power law amplitudes $\cP$ depending on which definition is taken.}.
Here, ${\bf l}_i = {\bf r}_{i+1}- {\bf r}_i$ denotes the
bond vector between two adjacent monomers $i$ and $i+1$
and $l^2 = \la {\bf l}^2_n \ra_n$ the mean-squared bond length. 
%
The bond-bond correlation function is generally believed to decrease {\em exponentially} 
\cite{FloryBook}. This belief is based on the few simple single chain models which have
been solved rigorously \cite{FloryBook,KP49} and on the assumption that {\em all}
long range interactions are negligible on distances larger than the screening length $\xi$.
Hence, only correlations along the backbone of the chains are expected to matter and it is
then straightforward to work out that an exponential cut-off is inevitable due to the
multiplicative loss of any information transferred recursively along the chain \cite{FloryBook}.

\begin{figure}[tb]
\includegraphics*[width=0.9\textwidth]{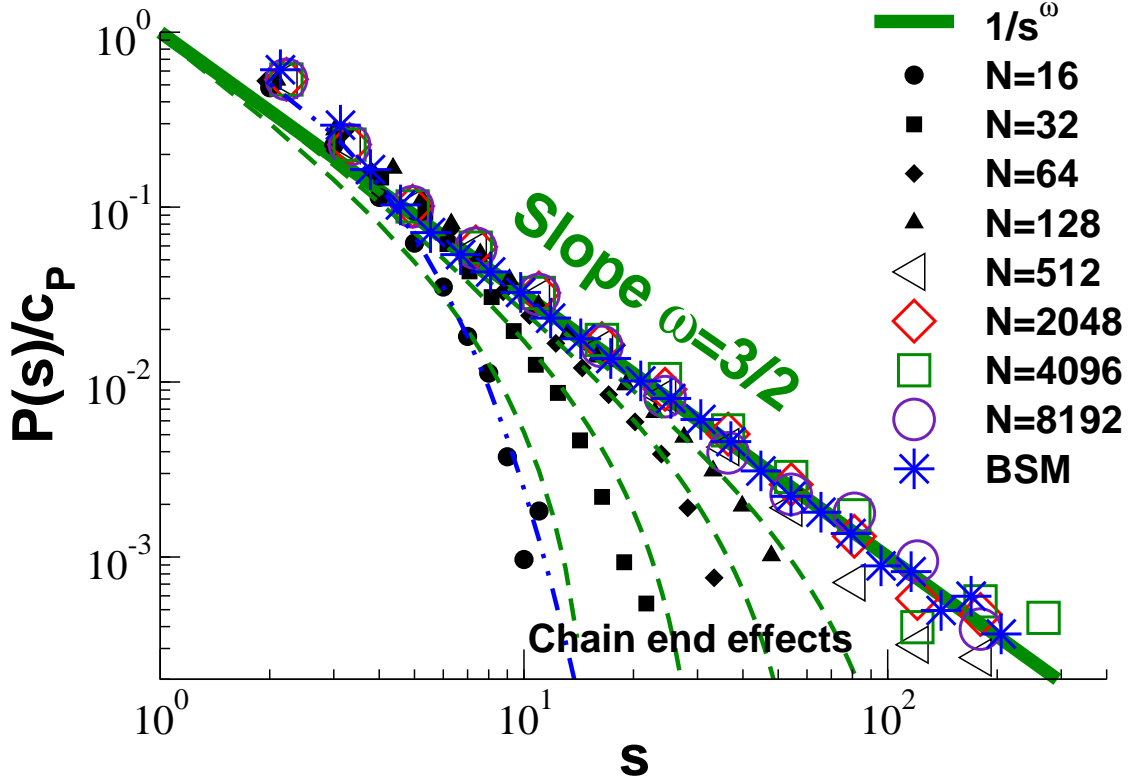}
\caption{(Color online)
The bond-bond correlation function $P(s)/\cP$ as a function of the 
curvilinear distance $s$. Various chain lengths are given for BFM. 
%
Provided that $1 \ll s \ll N$, all data sets collapse on the power law slope 
with exponent $\omega=3/2$ (bold line) as predicted by Eq.~(\ref{eq_P1s}). 
%
The dash-dotted curve $P(s) \approx \exp(-s/1.5)$ shows that exponential behavior is 
only compatible with very small chain lengths.
The dashed lines correspond to the theoretical prediction, Eq.~(\ref{eq_P1sN}), 
for short chains with $N=16,32,64$ and $128$ (from left to right). 
\label{fig_P1s}
}
\end{figure}

\paragraph*{Asymptotic behavior in the melt.}
That this reasoning must be {\em incorrect} follows immediately from the relation 
\begin{equation}
P(s) = \frac{1}{2l^2} \frac{d^2}{ds^2} \Rend^2(s)
\label{eq_RsP1sconnect}
\end{equation}
expressing the bond-bond correlation function as the curvature of the
second moment of the segment size distribution. 
It is obtained from the identity
$\la {\bf l}_n \cdot {\bf l}_m \ra = 
 \la \partial_n {\bf r}_n \cdot \partial_m {\bf r}_m \ra 
= - \partial_n \partial_m \la {\bf r}^2_{nm} \ra/2$.
(Note that the velocity correlation function is similarly related
to the second derivative of the mean-square displacement with respect to time
\cite{HansenBook}.)
Hence, $P(s)$ allows us to probe directly the non-Gaussian corrections
without any ideal contribution.
This relation together with Eq.~(\ref{eq_Rs_key}) suggests an algebraical decay 
$P(s) = \cP/s^{\omega}$ with
\begin{equation}
\omega = 3/2 
\ , \
\cP = c_1 \ (\be/l)^2 /8 
\approx \frac{1}{\rho l^2 \be} 
\label{eq_P1s}
\end{equation}
of the bond-bond correlation function for dense solutions and melts,
rather than the exponential cut-off expected from Flory's hypothesis.
This prediction (bold line) is perfectly confirmed by the larger chains 
($N > 256$) indicated in Fig.~\ref{fig_P1s}. 
In principle, the swelling coefficient, $c_1 \sim \cP$, may also be obtained from 
the power law amplitude of the bond-bond correlation function, however, to lesser 
accuracy than by the previous method (Fig.~\ref{fig_K1s}). One reason is that 
$P(s)$ decays very rapidly and does not allow a precise fit beyond $s \approx 10^2$. 
The values of $\cP$ obtained from $c_1$ are indicated in Tab.~\ref{tab_BFMBSMmelt}.
Data from the BSM have also been included in the figure to demonstrate the universality 
of the result. The vertical axis has been rescaled with $\cP$ which allows to 
collapse the data of both models.

\paragraph*{Finite chain-size corrections.}
\label{subsub_Ps_finiteN}
As can be seen for $N=16$, exponentials are compatible with the data of
short chains. This might explain how the power law scaling has been overlooked
in previous numerical studies, since good statistics for large chains ($N > 1000$)
has only become available recently. However, it is clearly shown that
$P(s)$ approaches systematically the scale free asymptote with increasing $N$.
The departure from this limit is fully accounted for by the theory if chain
end effects are carefully considered (dashed lines). Generalizing Eq.~(\ref{eq_P1s})
and using the Pad\'e approximation, Eq.~(\ref{eq_vPade}), perturbation theory yields
\begin{equation}
P(s) = \frac{\cP}{s^{3/2}} \frac{1+3x + 5x^2}{1+x} (1-x)^2
\label{eq_P1sN}
\end{equation} 
where we have set $x=\sqrt{s/N}$. For $x\ll 1$ this is consistent with Eq.~(\ref{eq_P1s}).
In the limit of large $s \rightarrow N$, the correlation functions vanish rigorously
as $P(s) \propto (1-x)^2$.
Considering that non-universal features cannot be neglected for short chain properties
and that the theory does not allow for any free fitting parameter, the agreement found 
in Fig.~\ref{fig_P1s} is rather satisfactory.


\subsection{Higher moments and associated coefficients}
\label{sub_p}

\paragraph*{Effective bond length and empirical swelling coefficients.}
The preceding discussion focused on the {\em second} moment of the segmental size 
distribution $G(r,s)$. 
We have also computed for both models higher moments $\la {\bf r}^{2p} \ra$ with $p \le 5$. 
If traced in log-linear coordinates as $y = (6^p p! \la {\bf r}^{2p} \ra / (2p+1)! s^p)^{1/p}$ 
{\em vs.} $x = s$ higher moments approach $\be^2$ from below --- just as the second moment 
presented in Fig.~\ref{fig_Rslog}. The deviations from ideality are now more pronounced 
and increase with $p$ (not shown).
The moments are compared in Fig.~\ref{fig_Kps} with Eq.~(\ref{eq_Rs_key})
where they are rescaled as $y=K_p(s)$ as defined in Eq.~(\ref{eq_Kpdef}) 
and plotted as functions of $x=\frac{3 (2^p p! p)^2}{2 (2p+1)!} \ \frac{\cp}{\sqrt{s}}$. 
The prediction is indicated by bold lines.
It is important that the {\em same} effective bond length $\be$ is obtained from 
the analysis of all functions $K_p(s)$ as illustrated in Fig.~\ref{fig_Kps}. 
Otherwise we would regard equilibration and statistics as insufficient.

The empirical swelling coefficients $\cp$ are obtained, as above in Sec.~\ref{sub_Rs},
by shifting the data horizontally. A good agreement with the expected $\cp/\ce \approx 1$ 
is found for both models and {\em all} moments as may be seen from  Tab.~\ref{tab_BFMBSMmelt}.
This confirms the renormalization of the Kuhn segment $b \rightarrow \be$ of the 
Gaussian reference chain in agreement with our discussion in Sec.~\ref{sub_theopert}.
Otherwise we would have measured empirical coefficients decreasing strongly as
$\cp/\ce \approx (b/\be)^{2p-3}$ with $p$. 
Since the effective bond length of non-interacting chains are known for
the BFM ($b \approx 2.688$) and the BSM ($b \approx 0.97$), one can simply check,
say for $p=5$, that the non-renormalized values would correspond to the ratios
$c_5/\ce \approx (2.688/3.244)^7 \approx 0.3$ for the BFM
and $c_5/\ce \approx (0.97/1.34)^7 \approx 0.1 $ for the BSM. 
This is clearly not consistent with our data.

It should be emphasized that both coefficients $\be$ and $\cp$ are more difficult to
determine for large $p$, since the linear regime for $x \ll 1$ in the representation 
chosen in Fig.~\ref{fig_Kps} becomes reduced. For large $x \gg 1$ one finds that
$y(x) \to 1$, i.e. $\la {\bf r}^{2p} \ra/\be^{2p}s^p \to 0$.
This trivial departure from both Gaussianity and the $1/\sqrt{s}$-deviations 
we try to describe, is due to the finite extensibility of chain segments of length $s$
which becomes more marked for larger moments probing larger segment sizes. 
The data collapse for both $x$-regimes is remarkable, however.
Incidentally, it should be noted that for the BSM the empirical swelling coefficients 
are slightly larger than expected. At present we do not have a satisfactory explanation 
for this altogether minor effect, but it might be attributed to the fact that neighbouring 
BSM beads along the chain strongly interpenetrate --- an effect not considered by the theory.

\paragraph*{Non-Gaussian parameter $\alphap$.}
The failure of Flory's hypothesis can also be demonstrated by means of the standard 
non-Gaussian parameter 
\begin{equation}
\alphap(s) \equiv 
1 - \frac{6^p p!}{(2p+1)!} \frac{\la {\bf r}_{nm}^{2p}\ra}{\la {\bf r}_{nm}^2 \ra^p}
\label{eq_alpha_p}
\end{equation} 
comparing the $2p$-th moment with the second moment ($p=1$).
In contrast to the closely related parameter $K_p(s)$ this has the advantage
that here two {\em measured} properties are compared without any tuneable parameter,
such as $\be$, which has to be fitted first. 
%
%
%
%
%
%
Fig.~\ref{fig_Kalphas} presents $\alphap(s)$ {\em vs.} $\ce/\sqrt{s}$
for the three moments with $p=2,3,4$. For each $p$ we find perfect data collapse 
for all chain lengths and both models and confirm the linear relationship 
$\alphap(s) \approx \Ustar(s)$ expected. 
The lines indicate the theoretical prediction
\begin{equation}
\alphap(s) = 
\left( \frac{3 \ (2^p p! p)^2}{2 \ (2p+1)!} - p \right) \frac{\ce}{\sqrt{s}}
\label{eq_alphap}
\end{equation}
which can be derived from Eq.~(\ref{eq_Rs_key}) by expanding the second 
moment in the denominator. An alternative derivation based on the coefficients
of the expansion of the generating function $G(q,s)$ in $q^2$ is indicated 
by Eq.~(\ref{eq_alphapAp}) in the Appendix. Having confirmed above that
$\cp/\ce \approx 1$, we assume in Eq.~(\ref{eq_alphap}) that $\cp = \ce$ 
to simplify the notation.
The prefactors $6/5$, $111/35$ and $604/105$ for $p=2$, $3$ and $4$ respectively
are nicely confirmed. They increase strongly with $p$, i.e. the non-Gaussianity 
becomes more pronounced for larger moments as already mentioned. 
%
%
Note also the curvature of the data at small $s$ due to the finite extensibility
of the segments which becomes more marked for higher moments. 
If one plots $\alphap(s)$ as a function of the {\em r.h.s.} of Eq.~(\ref{eq_alphap}) 
all data points for all moments and even for too small $s$ collapse on one master curve
(not shown) --- just as we have seen before in Fig.~(\ref{fig_Kps}).

\paragraph*{Correlations of different directions.}
A similar correlation function is presented in Fig.~\ref{fig_Kxys} which measures 
the non-Gaussian correlations of different spatial directions. It is defined by
\begin{equation}
K_{xy}(s) \equiv 1 - \frac{\la x^2 \ y^2 \ra}{\la x^2 \ra \la y^2 \ra}
\label{eq_Kxydef}
\end{equation}
for the two spatial components $x$ and $y$ of the vector ${\bf r}$
as illustrated by the sketch given at the bottom of Fig.~\ref{fig_Kxys}.
Symmetry allows to average over the three pairs of directions $(x,y)$, $(x,z)$ and $(x,z)$. 
Following the general scaling argument given in Sec.~\ref{sec_theo} we expect 
$K_{xy}(s) \approx \Ustar(s) \approx \ce/\sqrt{s}$ which is confirmed by the
perturbation result
\begin{equation}
K_{xy}(s) = \frac{6}{5} \frac{\ce}{\sqrt{s}} = K_2(s).
\label{eq_Kxy}
\end{equation}
This is nicely confirmed by the linear relationship found (bold line) 
on which all data from both simulation models collapse perfectly. 
The different directions of chain segments are therefore coupled. 
As explained in the Appendix (Eq.~(\ref{eq_KxyKalpha2})), $K_{xy}(s)$ and 
$\alpha_2(s)$ must be identical if the Fourier transformed segmental size 
distribution $G(q,s)$ can be expanded in terms of $q^2$ and this irrespective 
of the values the expansion coefficients take. 
Fig.~\ref{fig_Kxys} confirms, hence, that our computational systems are perfectly 
isotropic and tests the validity of the general analytical expansion.

The correlation function $K_{xy}$ is of particular interest since the 
zero-shear viscosity should be proportional to $\la \sigma_{xy}^2 \ra
\sim \la x^2 y^2 \ra = \la x^2 \ra \la y^2 \ra (1 - K_{xy}(s))$.
We assume here following Edwards \cite{DoiEdwardsBook} that only intrachain stresses 
contribute to the shear stress $\sigma_{xy}$.
Hence, our results suggest that the classical calculations \cite{DoiEdwardsBook} 
--- assuming incorrectly $K_{xy}=0$ ---
should be revisited.

\subsection{The segmental size distribution}
\label{sub_Grs}
%
%

We turn finally to the segmental size distribution $G(r,s)$ itself 
which is presented in Figs.~\ref{fig_Grs}, \ref{fig_dGrsbe} and \ref{fig_dGrsRes}. 
From the theoretical point of view $G(r,s)$ is the most fundamental property from 
which all others can be derived. It is presented last since it is computationally 
more demanding --- at least if high accuracy is needed --- and coefficients such as $\be$ 
may be best determined directly from the moments. 
The normalized histograms $G(r,s)$ are computed by counting the number of segment vectors 
between $r-dr/2$ and $r + dr/2$ with $dr$ being the width of the bin and one
divides then by the spherical bin volume. Since the BFM model is a lattice model, 
this volume is not $4\pi r^2 dr$ but given by the number of lattice sites the 
segment vector can actually point to for being allocated to the bin. 
Incorrect histograms are obtained for small $r$ if this is not taken into account. 
(Averages are taken over all segments and chains, just as before.) 
Clearly, non-universal physics must show up for small vector length $r$ and small 
curvilinear distance $s$ and we concentrate therefore on values $r \gg \sigma$ and $s \ge 31$.

When plotted in linear coordinates as in Fig.~\ref{fig_Grs}, $G(r,s)$ compares roughly 
with the Gaussian prediction $G_0(r,s)$ given by Eq.~(\ref{eq_G0rs}), but presents a 
distinct depletion for small segment sizes with $n \equiv r/ \be \sqrt{s} \ll 1$
and an enhanced regime for $n \approx 1$. A second depletion region for large $n \gg 1$ 
--- expected from the finite extensibility of the segments --- can be best seen in the 
log-log representation of the data (not shown). 
To analyse the data it is better to consider instead of $G(r,s)$ the relative deviation
$\delta G(r,s) / G_0(r,s) = G(r,s)/G_0(r,s) -1$ which should further be divided by the 
strength of the segmental correlation hole, $\ce/\sqrt{s}$. As presented in Fig.~\ref{fig_dGrsbe} 
this yields a direct test of the key relation Eq.~(\ref{eq_dGrsbe}) announced in the Introduction.
The figure demonstrates nicely the scaling of the data for all $s$ and for both models.
It shows further a good collapse of the data close to the universal function $f(n)$ 
predicted by theory (bold line). 
Note that the depletion scales as $1/n$ for small segment sizes (dashed line).
The agreement of simulation and theory is by all standards remarkable. 
(Obviously, error bars increase strongly for $n \gg 1$ where $G_0(r,s)$ decreases strongly.
The regime for very large $n$ where the finite extensibility of segments matters has been 
omitted for clarity.)
We emphasize that this scaling plot depends very strongly on the value $\be$  
which is used to calculate the Gaussian reference distribution.

If a precise value is not available we recommend to use instead the scaling variable 
$m = r/\Rend(s)$ for the horizontal axis, i.e. to replace the scale $\be \sqrt{s}$ 
estimated from the behaviour of asymptotically long chains by the measured (mean-squared) 
segment size for the given $s$. The Gaussian reference distribution is then accordingly 
$G_0(m,\Rend(s)) = (3/2\pi\Rend(s)^2) \exp(- \frac{3}{2} m^2 )$. 
The corresponding scaling plot is given in Fig.~\ref{fig_dGrsRes}.
It is similar and of comparable quality as the previous plot. 
Changing the scaling variable from $n=r/\be \sqrt{s}$ to
$m = r/\Rend(s) \approx (r/\be \sqrt{s}) (1 + \ce/2\sqrt{s})$
changes somewhat the universal function. Expanding the previous
result, Eq.~(\ref{eq_dGrsbe}), this adds even powers of $m$ to the 
function $f(n)$ given in Eq.~(\ref{eq_dGrsbe}) 
\begin{equation}
f(n) \Rightarrow f(m) = \sqrt{\frac{3\pi}{32}} \left( 
-\sqrt{\frac{24}{\pi}} -\frac{2}{m} + 9 m + \sqrt{\frac{24}{\pi}} m^2 - \frac{9}{2} m^3 
\right).
\label{eq_dGrsRe}
\end{equation}
That the two additional terms in the function are correct can be seen 
by computing the second moment $4\pi \int dr r^4 \delta G(r,s)$
which must vanish by construction. The rescaled relative deviation
is somewhat broader than in the previous plot due to the additional
term scaling as $m^2$.
As already stressed this scaling does not rely on the effective
bond length $\be$ and is therefore more robust.
It has the nice feature that it underlines that there is only {\em one}
characteristic length scale relevant for the swelling induced by the
segmental correlation hole, the typical size of the chain segment itself.

\section{Conclusion \label{sec_conc}}

%
%
\paragraph*{Issues covered and central theoretical claims.}
We have revisited Flory's famous ideality hypothesis for long polymers in the melt 
by analyzing both analytically and numerically the segmental size distribution $G(r,s)$ 
and its moments for chain segments of curvilinear length $s$.
We have first identified the general mechanism that gives rise to deviations from 
ideal chain behavior in dense polymer solutions and melts (Sec.~\ref{sec_theo}). 
This mechanism rests upon the interplay of chain connectivity and the incompressibility 
of the system which generates an effective repulsion between chain segments 
(Fig.~\ref{fig_corehole}). This repulsion scales like $\Ustar(s) \approx \ce/\sqrt{s}$ 
where the ``swelling coefficient" $\ce \approx 1/\be^3\rho$ sets the strength of the interaction.
It is strong for small segment length $s$, but becomes weak for $s\rightarrow N$ 
in the large-$N$ limit. The overall size of a long chain thus remains almost `ideal', 
whereas subchains are swollen as described by Eq.~(\ref{eq_Rs_key}). 
Most notably, the relative deviation $\delta G(r,s)/G_0(r,s)$ of the segmental size 
distribution from Gaussianity should be proportional to $\Ustar(s)$. 
As a function of segment size $r$, the repulsion manifests itself by a
strong $1/r$-depletion at short distances $r \ll \be \sqrt{s}$ and a
subsequent shift of the histogram to larger distances (Eq.~(\ref{eq_dGrsbe})).  

\paragraph*{Summary of computational results.}
Using Monte Carlo and molecular dynamics simulation of two coarse-grained 
polymer models we have verified numerically the theoretical predictions for long
and flexible polymers in the bulk.
We have explicitly checked (e.g., Figs.~\ref{fig_Krecurs}, \ref{fig_Kalphas}, \ref{fig_dGrsRes}) 
that the relative deviations from Flory's hypothesis scale indeed as $1/\sqrt{s}$. Especially, 
the measurement of the bond-bond correlation function $P(s)$, being the second derivative of 
the second moment of $G(r,s)$ with respect of $s$, allows a very precise verification 
(Fig.~\ref{fig_P1s}) and shows that higher order corrections beyond the first-order perturbation 
approximation must be small.
The most central and highly non-trivial numerical verification concerns the data collapse 
presented in Figs.~\ref{fig_dGrsbe} and \ref{fig_dGrsRes} for the segmental size distribution 
of both computational models. All other statements made in this paper can be derived and 
understood from this key finding. It shows especially that the swelling coefficient $\ce$
must be close to the predicted value, Eq.~(\ref{eq_cpce}).

It is well known \cite{ME82} that the effective bond length is difficult to 
predict at low compressibility and no attempt has been done to do so in this paper. 
We show instead how the systematic swelling of chain segments -- once understood -- 
may be used to extrapolate for the effective bond length of {\em asymptotically} 
long chains. Figs.~\ref{fig_K1s} and \ref{fig_Kps} indicate how this may be done
using Eq.~(\ref{eq_Rs_key}). The high precision of our data is demonstrated 
in Fig.~\ref{fig_dGrsbe} by the successful scaling of the segmental size distribution.

For several moments $\la {\bf r}^{2p} \ra$ we have also fitted empirical swelling 
coefficients $\cp$ using Eq.~(\ref{eq_Rs_key}). In contrast to the effective bond length $\be$ 
these coefficients are rather well predicted by one-loop perturbation theory 
{\em if} the bond length $b$ of the reference Hamiltonian is {\em renormalized}
to the effective bond length $\be$, as we have conjectured in Sec.~\ref{subsub_brenorm}. 
Since the empirical swelling coefficients, $\cp \approx \ce (b/\be)^{2p-3}$,
would otherwise strongly depend on the moment taken, as shown in Eq.~(\ref{eq_Rppert}), 
our numerical data (Tab.~\ref{tab_BFMBSMmelt}) clearly imply $b /\be \approx 1$.
Minor deviations found for the BSM samples may be attributed to the fact that monomers 
along the BSM chains do strongly overlap --- an effect not taken into account by the theory.
To clarify ultimately this issue we are currently performing a numerical study 
where we systematically vary both the compressibility and the bond length of the BSM.

%
%
%
\paragraph*{General background and outlook.}
The most striking result presented in this work concerns the power law decay found for 
the bond-bond correlation function, $P(s) \propto 1/s^{3/2}$ (Fig.~\ref{fig_P1s}). 
This result suggests an analogy with the well-known long-range velocity correlations 
found in dense fluids by Alder and Wainwright nearly fourty years ago
\cite{Alder70,HansenBook}.
In both cases, the ideal uncorrelated object is a random walker which is weakly perturbed 
(for $d > 2$) by the self-interactions generated by global constraints. Although these 
constraints are different (momentum conservation for the fluid, incompressibility 
for polymer melts) the weight with which these constraints increase the stiffness of the 
random walker is always proportional to the return probability.
It can be shown that the correspondence of both problems is
mathematically rigorous if the fluid dynamics is described on the level of 
the {\em linearized} Navier-Stokes equations \cite{papPs}.
%
%

We point out that the physical mechanism which has been sketched above is rather 
general and should not be altered by details such as a finite persistence length 
--- at least not as long as nematic ordering remains negligible and the polymer chains 
are sufficiently {\em long}. 
(Similarly, velocity correlations in dense liquids must show an analytical decay 
for sufficiently {\em large} times irrespective of the particle mass and the local
static structure of the solution.)
While this paper focused exclusively on scales beyond the correlation length of the 
density fluctuations, i.e.  $q \xi \ll 1$ or $s/g \gg 1$, where the polymer solution 
appears incompressible, effects of finite density and compressibility can be readily 
described within the same theoretical framework and will be presented elsewhere \cite{papPs}. 
To test our predictions, {\em flexible} chains should be studied preferentially,
since the chain length required for a clear-cut description increases strongly with 
persistence length. This is in fact confirmed by preliminary and on-going simulations 
using the BSM algorithm. 

In this work we have only discussed properties in real space as a function of 
the curvilinear distance $s$. These quantities are straightforward to compute in a 
computer simulation but are barely experimentally relevant. The non-Gaussian deviations 
induced by the segmental correlation hole arise, however, also for an experimentally 
accessible property, the intramolecular form factor (single chain scattering function)
$F(q)$. As explained at the end of the Appendix, the form factor can be readily obtained 
by integrating the Fourier transformed segmental size distribution given in Eq.~(\ref{eq_dGrsbe}). 
This yields
\begin{equation}
q^2 F(q) \approx \frac{12}{\be^2} \left( 1 - \frac{3}{8} \frac{\be q}{\be^3\rho} \right)
\label{eq_Fq}
\end{equation}
in agreement with the result obtained in Refs.~\cite{BJSOBW07,WBJSOMB07}
by direct calculation of the form factor for very long equilibrium polymers. 
As a consequence of this, the Kratky plot ($q^2F(q)$ {\em vs.} wave vector $q$) should
not exhibit the plateau expected for Gaussian chains in the scale-free regime,
but rather noticeable non-monotonic deviations. See Fig.~3 of \cite{WBJSOMB07}. 
This result suggests to revisit experimentally this old pivotal problem of polymer science.
%

Our work is part of a broader attempt to describe systematically the effects of
correlated density fluctuations in dense polymer systems, both for static
\cite{ANS96,ANS03,ANS05a,ANS05b} and dynamical \cite{ANS97,ANS98,MWBBL03} properties.  
An important unresolved question is for instance whether the predicted long-range 
{\em repulsive} forces of van der Waals type (``Anti-Casimir effect") \cite{ANS05a,ANS05b}
are observable, for instance in the oscillatory decay of the standard density pair-correlation 
function of dense polymer solutions.
Since the results presented here challenge an important concept of polymer physics, 
they should hopefully be useful for a broad range of theoretical approaches which 
commonly assume the validity of the Gaussian chain model down to molecular scales 
\cite{CSGK91,Fuchs,Schulz}. 
This study shows that a polymer in dense solutions should not be viewed as {\em one} 
soft sphere (or ellipsoid) \cite{Likos01,EM01,Guenza04}, but as a hierarchy of nested segmental 
correlation holes of all sizes aligned and correlated along the chain backbone 
(Fig.~\ref{fig_corehole} (b)). 
We note that similar deviations from Flory's hypothesis have been reported recently for
linear polymers \cite{CSGK91,SMK00,Auhl03} and polymer gels and networks \cite{Sommer05,SGE05}.
The repulsive interactions should also influence the polymer dynamics, since
strong deviations from Gaussianity are expected on the scale where entanglements
become important, hence, quantitative predictions for the entanglement length $\Ne$
have to be regarded with more care. 
The demonstrated swelling of chains should be included in the popular primitive path 
analysis for obtaining $\Ne$ \cite{EveraersScience}, especially if `short' chains ($N < 500$) 
are considered. The effect could be responsible for observed deviations from Rouse behavior 
\cite{Paul91a,PS04} as may be seen by considering the correlation function 
$C_{pq} \equiv \la {\bf X}_p \cdot {\bf X}_q \ra$ of the Rouse modes 
${\bf X}_p = \frac{1}{N} \int_{n=0}^N dn {\bf r}_n \cos( n p \pi/N)$
where $p,q = 0,\ldots,N-1$ \cite{Verdier66,DoiEdwardsBook}. 
Using $({\bf r}_n - {\bf r}_m)^2 = {\bf r}_n^2 + {\bf r}_m^2 - 2 {\bf r}_n \cdot {\bf r}_m$
for the segment size, this correlation function can be readily expressed as an integral over 
the second moment of the segmental size distribution
\begin{equation}
C_{pq}  
= - \frac{1}{2N^2} \int_0^N dn \int_0^N dm \la ({\bf r}_n - {\bf r}_m)^2 \ra
\cos(n p \pi/N) \cos(m p \pi/N)
\label{eq_Cpq}
\end{equation}
which can be solved using our result Eq.~(\ref{eq_Rs_key}).
This implies for instance for $p=q$ that
\begin{equation}
C_{pp} = \frac{N \be^2}{2 (\pi p)^2} 
\left(1 - \frac{\pi}{\sqrt{8}} \frac{c_1}{\sqrt{N/p}} \right).
\label{eq_Cpqcorrection}
\end{equation}
The bracket entails an important correction with respect to the classical
description given by the prefactor \cite{DoiEdwardsBook}. 
We are currently working out how static corrections, such as those for $C_{pp}$,
may influence the dynamics for polymer chains {\em without} topological constraints.
(This may be realized, e.g., within the BFM algorithm by using the L26 moves 
described in Sec.~\ref{sub_BFM}.)

Moreover, for thin polymer films of width $H$ the repulsive interactions are known to be 
stronger than in the bulk \cite{ANS03}. This provides a mechanism to rationalize the trend
towards swelling observed experimentally \cite{JKHBR99} and confirmed computationally 
\cite{CMWJB05}: 
\begin{equation}
1- \frac{R_{x}^2(s)}{b_{x}^2 s} = \log(s)/H.
\label{eq_Rs_slit}
\end{equation}
(Prefactors omitted for clarity.)
Here $R_{x}(s)$ and $b_{x}$ denote the components of the segment size and the 
effective bond length parallel to the film.
It also explains the (at first sight surprising) systematic 
{\em increase} of the polymer dynamics with decreasing film thickness \cite{MKCWB07}. 
Specifically, the parallel component of the monomer mean-squared displacement $g_x(t)$
is expected to scale as
%
$g_x(t) \approx R_{x}^2(s(t)) \propto t^{1/4} (1 + \log(t)/H)$
for long reptating chains where $s(t) \propto t^{1/4}$ \cite{DoiEdwardsBook}.
(The corresponding effect for the three-dimensional bulk should be small, however.)
For the same reason (flexible) polymer chains close to container walls must be more 
swollen and, hence, {\em faster} on intermediate time scales than their peers in the bulk. 

\begin{acknowledgments}
We thank T.~Kreer, S.~Peter and A.N.~Semenov (all ICS, Strasbourg, France),
S.P.~Obukhov (Gainesville, Florida) and M.~M\"uller (G\"ottingen, Germany)
for helpful discussions.
A generous grant of computer time by the IDRIS (Orsay) is also gratefully acknowledged. 
J.B. acknowledges financial support by the IUF and from the European Community's
``Marie-Curie Actions" under contract MRTN-CT-2004-504052.
\end{acknowledgments}

\appendix
\section{Moments of the segmental size distribution and their generating function}
\label{app_pert}

\begin{figure}[tb]
\includegraphics*[width=0.9\textwidth]{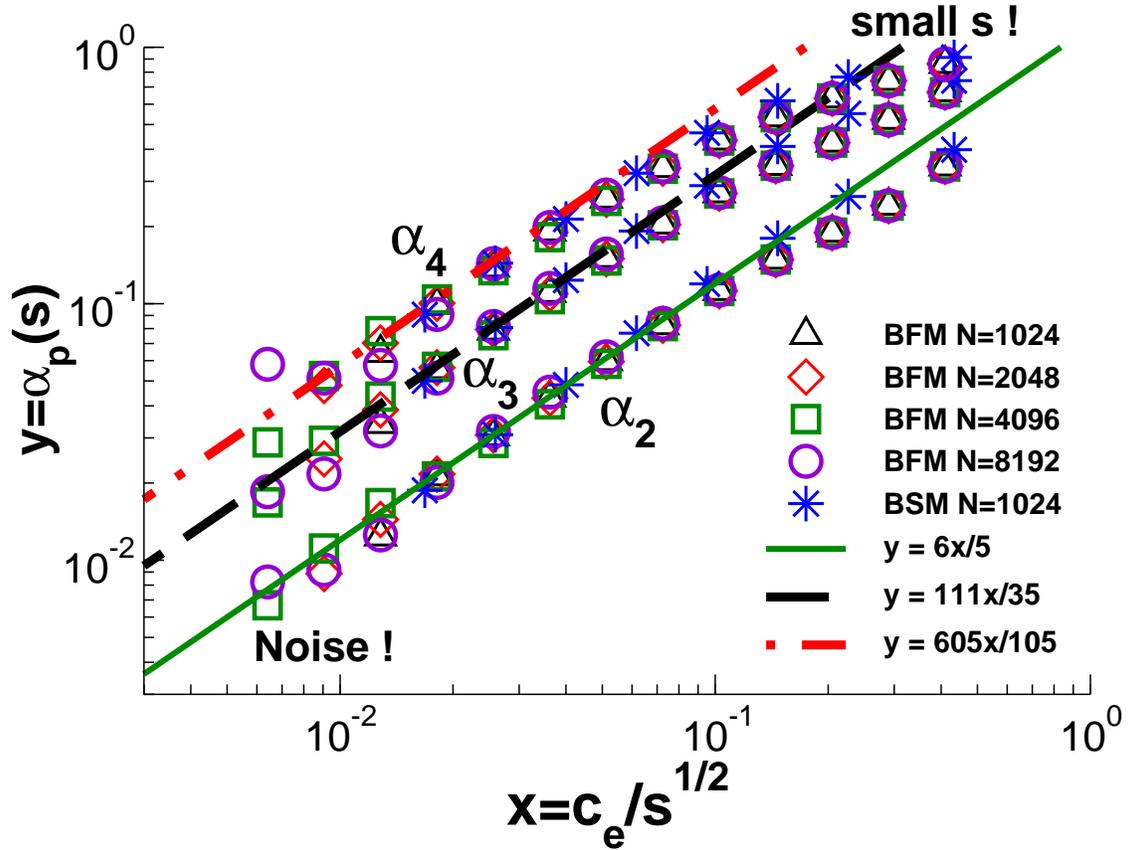}
\caption{(Color online)
Non-Gaussian parameter $\alphap(s)$ computed for the end-to-end 
distance of chain segments as a function of $\ce/\sqrt{s}$. 
Perfect data collapse for all chain lengths and both simulation models is obtained for each $p$.
A linear relationship over nearly two orders of magnitude is found as theoretically expected.
Data for three moments ($p=2,3,4$) are indicated showing a systematic {\em increase} 
of non-Gaussianity with $p$. 
The data curvature for small $s$ becomes more pronounced for larger $p$.
\label{fig_Kalphas}
}
\end{figure}

\begin{figure}[tb]
\includegraphics*[width=0.9\textwidth]{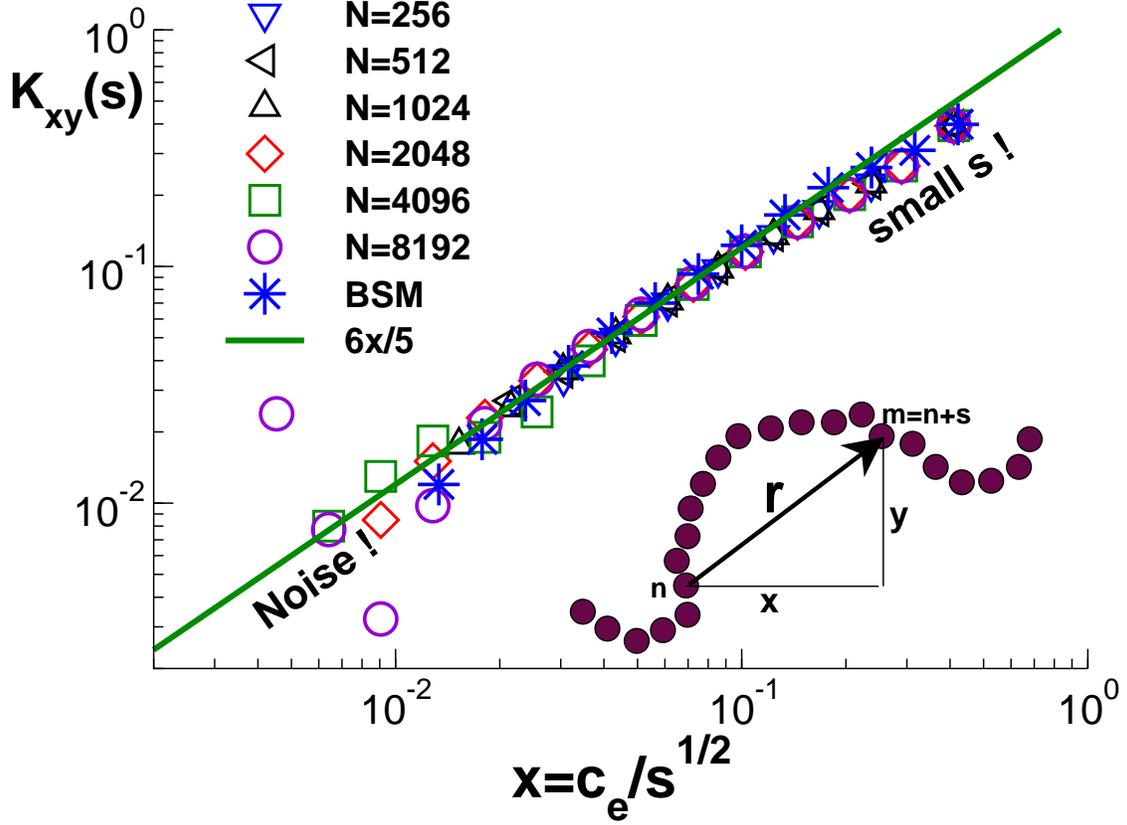}
\caption{(Color online)
Plot of $K_{xy}(s) = 1 - \la x^2 y^2 \ra / \la x^2 \ra \la y^2 \ra$ 
averaged over all pairs of monomers ($n,m=n+s$) and three different direction pairs
as a function of $\ce/\sqrt{s}$. 
As indicated by the sketch at the bottom of the figure, $K_{xy}(s)$ measures the 
correlation of the components of the segment vector ${\bf r}$.
All data points collapse and show again a linear relationship $K_{xy} \approx \Ustar(s)$.
Different directions are therefore coupled!
No curvature is observed over two orders of magnitude confirming that
higher order perturbation corrections are negligible.
Noise cannot be neglected for large $s > 100$ and finite segment-size effects 
are visible for $s \approx 1$.
\label{fig_Kxys}
}
\end{figure}

\begin{figure}[t]
\centering
\includegraphics*[width=0.9\textwidth]{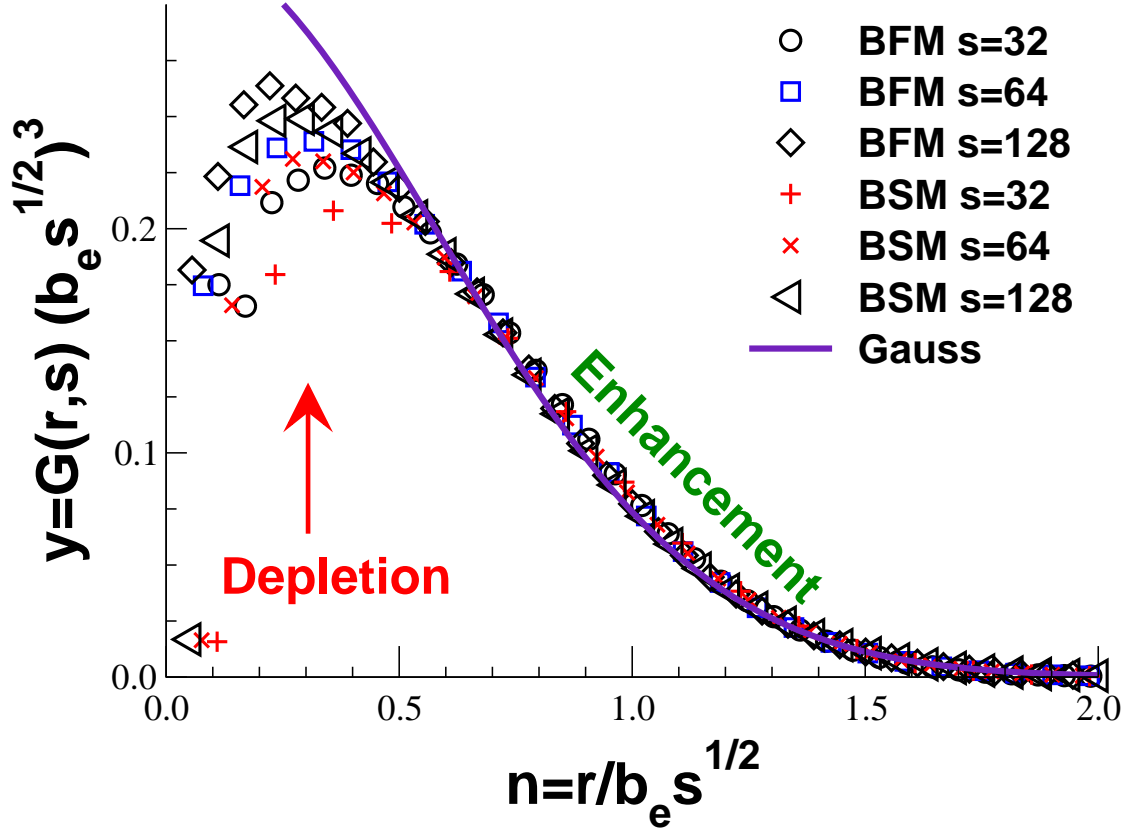}
\caption{(Color online)
Segment size distribution $y=G(r,s) (\be s^{1/2})^3$
{\em vs.} $n=r/\be s^{1/2}$ for several $s$ as indicated in the figure.
Only data for BFM with $N=2048$ and BSM with $N=1024$ are presented.
(A similar plot can be achieved by renormalizing the axes using $\Rend(s)$ 
instead of $\be s^{1/2}$).
The bold line denotes the Gaussian behaviour $y=(3/2\pi)^{3/2} \exp(-3n^2/2)$.
One sees that compared to this reference the measured distributions are depleted
for small $n \ll 1$ (where the data does not scale) and enhanced for $n \approx 1$.
\label{fig_Grs}
}
\end{figure}

\begin{figure}[t]
\centering
\includegraphics*[width=0.9\textwidth]{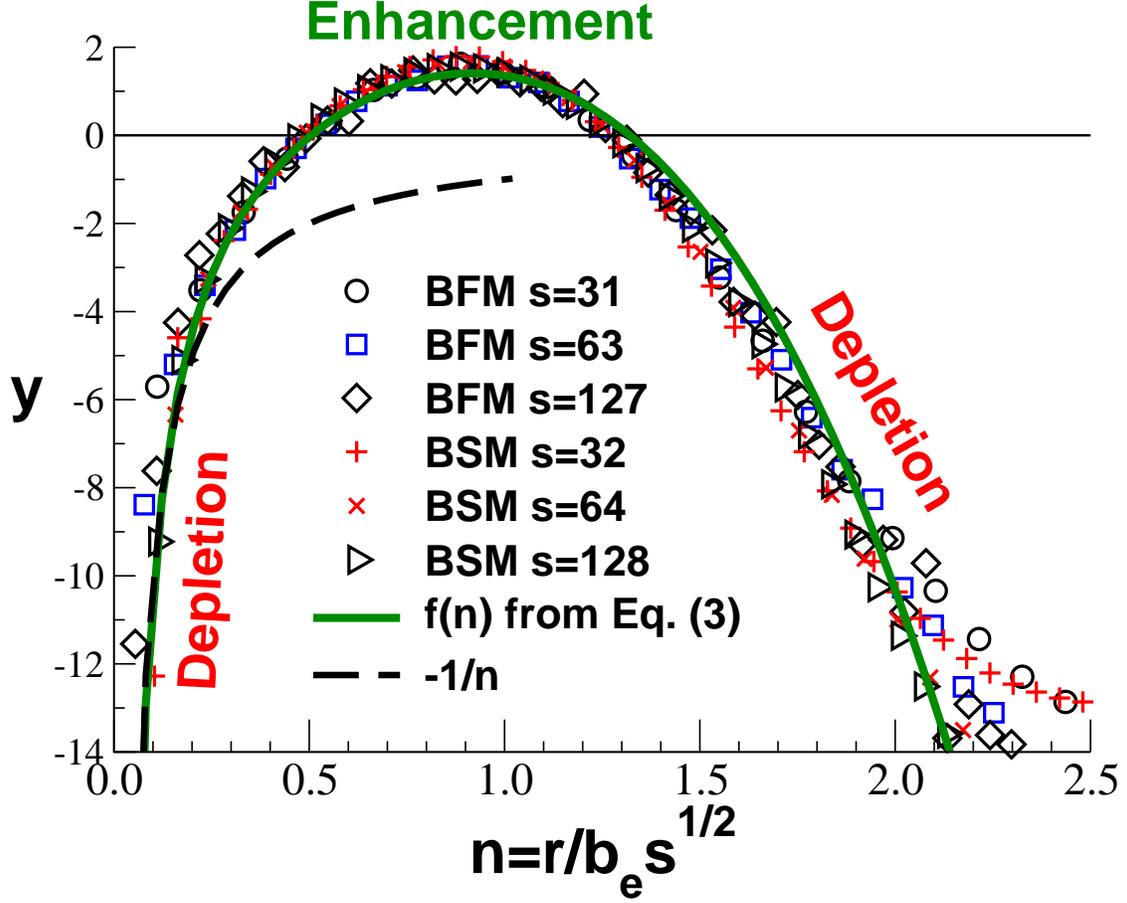}
\caption{(Color online)
Deviation $\delta G(r,s) = G(r,s) - G_0(r,s)$
of the measured segmental size distribution 
from the Gaussian behavior $G_0(r,s)$ expected from Flory's hypothesis
for several $s$ and both models as indicated in the figure.
As suggested by Eq.~(\ref{eq_dGrsbe}), we have plotted
$y= (\delta G(r,s)/G_0(r,s))/(\ce/\sqrt{s})$ as a function of $n=r/\be \sqrt{s}$.
The Gaussian reference distribution has been computed according to Eq.~(\ref{eq_G0rs})
for the measured effective bond length $\be$. 
A close to perfect data collapse is found for both models. 
This shows that the deviation scales linearly with $\Ustar(s) \approx \ce/\sqrt{s}$,
as expected. The bold line indicates the universal function of $f(n)$ predicted by 
Eq.~(\ref{eq_dGrsbe}). 
\label{fig_dGrsbe} 
}
\end{figure}

\begin{figure}[t]
\centering
\includegraphics*[width=0.9\textwidth]{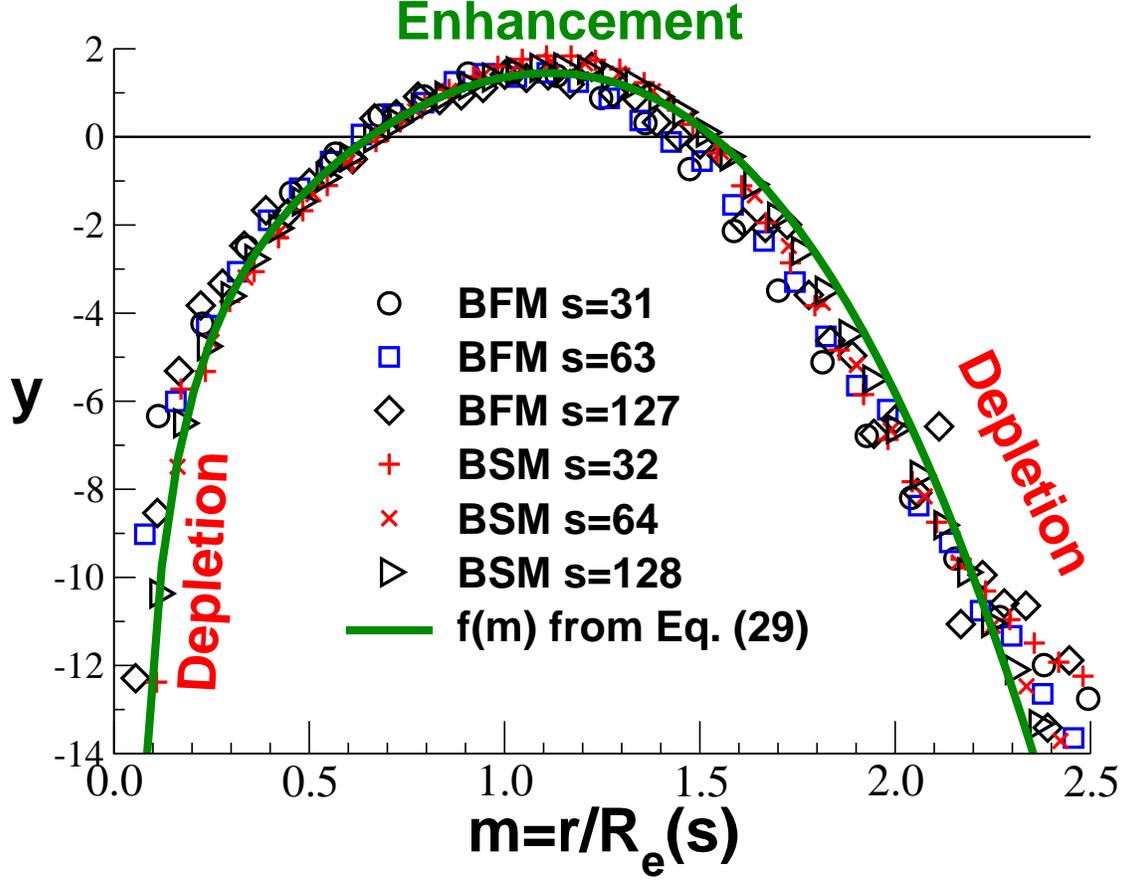}
\caption{(Color online)
Replot of the relative deviation of the measured segment size distribution,
$y = (\delta G(r,s)/G_0(r,s)))/(\ce/\sqrt{s})$, as a function of $m=r/\Rend(s)$.
The figure highlights that the measured segment size is 
the only length scale relevant for describing the deviation from Flory's hypothesis.
The same data sets and symbols are used as in the previous Fig.~\ref{fig_dGrsbe}.
\label{fig_dGrsRes} 
}
\end{figure}

Higher moments of the segmental size distribution $G(r,s)$ can be systematically 
obtained from its Fourier transformation
\begin{equation}
\nonumber
G(q,s)=\int d^3{\bf r} \ G(r,s) \ \exp(i{\bf q \cdot r}), 
\label{eq_Gqpr}
\end{equation}
which is in this context sometimes called the ``generating function" \cite{KampenBook}.
For an ideal Gaussian chain, the generating function is then $G_0(q,s)=\exp(-s q^2a^2)$
where we have used $a^2 = b^2/6$ instead of the bond length $b^2$ to simplify the notation.
Moments of the size distribution are given by proper derivatives of $G(q,s)$ taken at $q=0$.
For example, $\langle {\bf r}^{2p}\rangle=(-1)^p\Delta^p G(q,s)_{q=0}$
(with $\Delta$ being the Laplace operator with respect to the wave vector ${\bf q}$).
A moment of order $2p$ is, hence, linked to only {\em one} coefficient $A_{2p}$ in 
the systematic expansion, $G(q,s) = \sum_{p=0} A_{2p} q^{2p}$, of $G(q,s)$ around $q=0$.
For our example this implies 
\begin{equation}
\la {\bf r}^{2p} \ra = (-1)^p (2p+1)! \ A_{2p}
\label{eq_r2pA2p}
\end{equation}
in general and more specifically for a Gaussian distribution
${\langle {\bf r}^{2p}\rangle}_0=\frac{(2p+1)! }{p!}s^p a^{2p}$. 
The non-Gaussian parameters read, hence,
\begin{equation}
\alphap(s) \equiv 
1 - \frac{6^p p!}{(2p+1)!} \frac{\langle {\bf r}^{2p} \rangle}{\langle {\bf r}^2 \rangle^p}
= 1 - p! \ \frac{A_{2p}}{A_2^p},
\label{eq_alphapAp}
\end{equation}
which implies (by construction) $\alphap = 0$ for a Gaussian distribution.
As various moments of the same global order $2p$ are linked to the same $A_{2p}$ they differ 
by a multiplicative constant independent of the details of the (isotropic) distribution $G(q,s)$. 
For example, $\langle {\bf r}^{2}\rangle= 6 |A_2|$, 
$\langle {\bf r}^{4}\rangle=120 A_4$, 
$\langle x^{2}\rangle = \langle y^{2}\rangle = 2 |A_2|$, 
$\langle x^{2}y^{2}\rangle=8A_4$
with $x$ and $y$ denoting the spatial components of the segment vector ${\bf r}$.
Using Eq.~(\ref{eq_alphapAp}) for $p=2$ it follows that
\begin{equation}
K_{xy}(s) \equiv 1 - \frac{\langle x^{2}y^{2}\rangle}{\langle x^{2}\rangle\langle y^{2}\rangle}
= 1- 2 \frac{A_4}{A_2^2} = \alpha_2(s),
\label{eq_KxyKalpha2}
\end{equation}
i.e. the properties $\alpha_2(s)$ and $K_{xy}(s)$ discussed in Figs.~\ref{fig_Kalphas} and 
\ref{fig_Kxys} must be identical in general provided that 
$G(q,s)$ is isotropic and can be expanded in $q^2$.

\begin{figure}[tb]
\includegraphics*[width=0.6\textwidth]{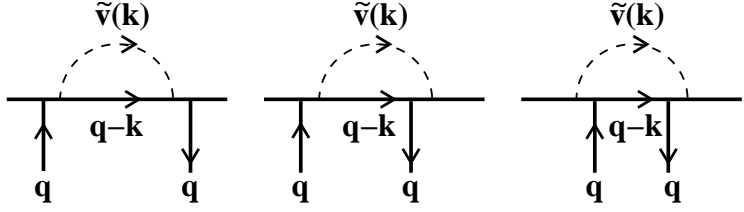}
\caption{
Interaction diagrams used in reciprocal space for the calculation of $\delta G(q,t)$ 
in the scale free limit.
There exist three nonzero contributions to first-order perturbation, 
the first involving two points inside the segment (first two lines of Eq.~(\ref{eq_deltaGqt})), 
the second one point inside and one outside the segment
(third line of Eq.~(\ref{eq_deltaGqt})) and the third one point on either side 
of the segment (last line of Eq.~(\ref{eq_deltaGqt})). 
Momentum $q$ flows from one correlated point to the other. 
Integrals are performed over the momentum $k$.
Dotted lines denote the effective interactions $\veff(k)$ given by Eq.~(\ref{eq_v1q}),
bold lines the propagators which carry each a momentum $q$ or $q-k$ as indicated.
\label{fig_diagGqt}
}
\end{figure}

We turn now to specific properties of $G(q,s)$ computed for formally {\em infinite} polymer chains 
in the melt. In practice, these results are also relevant for small segments in large chains, 
$N \gg s \gg 1$, and, especially, for segments located far from the chain ends.
These chains are nearly Gaussian and the generating function can be written as 
$G(q,s)=G_0(q,s)+\delta G(q,s)$ where 
$\delta G(q,s) = - \langle U G \rangle_0 + \langle U \rangle_0 \langle G \rangle_0$ 
is a small perturbation under the effective interaction potential $\veff(q)$ given by Eq.~(\ref{eq_v1q}). 
To compute the different integrals it is more convenient to work in Fourier-Laplace space ($q,t$) 
with $t$ being the  Laplace variable conjugate to $s$:
\begin{equation}
\nonumber
\delta G(q,t) = \int_0^{\infty} ds \ \delta G(q,s) e^{-s t}.
\end{equation}
As illustrated in Fig.~\ref{fig_diagGqt}, there a three contributions to this perturbation: 
one due to interactions between two monomers inside the segment (left panel), 
one due to interactions between an internal monomer and an external one (middle panel) and
one due to interactions between two external monomers located on opposite sides (right panel). 
In analogy to the derivation of the form factor described in Ref.~\cite{BJSOBW07} this yields:
\begin{eqnarray}
\delta G(q,t) & = & \nonumber
-\frac{1}{(q^2a^2 + t)^2}\frac{v}{4\pi a^3}\left(\sqrt{q^2a^2 + t} - \sqrt{t}\right)  \\ \nonumber
& + & \frac{1}{(q^2a^2 + t)^2}\frac{v}{4\pi q a^2\xi^2}
\left(\mbox {Arctan}\left[\frac{q a}{a / \xi +\sqrt{t}}\right] - 
\frac{q a}{a/ \xi+\sqrt{q^2a^2 + t}}\right)\\ \nonumber
&-&\frac{1}{q^2a^2 + t}\frac{2v}{ 4\pi q a^4}
\left(\mbox{Arctan}\left[\frac{q a}{a / \xi + \sqrt{t}}\right] - 
\frac{q a}{a / \xi + \sqrt{q^2a^2 + t}}\right)\\ 
& - &\frac{v\xi^2}{4\pi q a^6}
\left(\mbox{Arctan}\left[\frac{q a}{\sqrt{t}}\right] - \frac{q a}{\sqrt{q^2a^2 + t}}\right).
\label{eq_deltaGqt}
\end{eqnarray}
The graph given in the left panel of Fig.~\ref{fig_diagGqt} corresponds to the first two lines,
the middle panel to the third line and the right panel to the last one.
Seeking for the moments we expand $\delta G(q,t)$ around $q=0$. 
Having in mind chain strands counting many monomers ($s\gg 1$), 
we need only to retain the most singular terms for $t \rightarrow 0$.
Defining the two dimensionless constants 
$d = v \xi / 3 \pi a^4  = 12 v \xi / \pi b^4$ and
$c = (3 \pi^{3/2} a^3 \rho )^{-1} = \sqrt{24/\pi^3}/b^3 \rho$ this expansion
can be written as
\begin{eqnarray}
\label{eq_Gexpansion}
\delta G(q,t) = 
& - & \frac{1}{1!} \frac{\Gamma(2)}{t^2}  \ d         \ a^2 q^2 
  +   \frac{1}{1!} \frac{\Gamma(3/2)}{t^{3/2}} \  c  \ a^2 q^2  + \ldots 
\\ \nonumber
& + & \frac{2}{2!} \frac{\Gamma(3)}{t^3} \  d         \ a^4 q^4 
  -   \frac{1}{2!} \frac{16}{5} \frac{\Gamma(5/2)}{t^{5/2}}  \  c   \  a^4 q^4 + \ldots  
\\ \nonumber
& - & \frac{3}{3!} \frac{\Gamma(4)}{t^4} \  d  \  a^6 q^6 
  +   \frac{1}{3!} \frac{216}{35} \frac{\Gamma(7/2)}{t^{7/2}} \  c  \   a^6 q^6 + \ldots 
\\ \nonumber
& + & \ldots 
\end{eqnarray}
where we have used Euler's Gamma function $\Gamma(\alpha)$ \cite{abramowitz}.
The first leading term at each order in $q^2$ --- being proportional to the coefficient $d$ 
--- ensures the renormalization of the effective bond length. The next term scaling 
with the coefficient $c$ corresponds to the leading finite strand size correction.
Performing the inverse Laplace transformation 
$\Gamma(\alpha)/t^{\alpha} \rightarrow s^{\alpha-1}$
and adding the Gaussian reference distribution $G_0(q,s)$
this yields the $A_{2p}$-coefficients for the expansion of $G(q,s)$
around $q=0$:
\begin{eqnarray}
\nonumber
A_0 & = & 1 \\ \nonumber
A_2 & = & - a^2 s \left(1 + d - \frac{c}{\sqrt{s}}  \right) 
\\ \nonumber
A_4 & = & \frac{1}{2} a^4 s^2 \left(1 + 2 d - \frac{16}{5} \frac{c}{\sqrt{s}} \right) 
\\ \nonumber 
A_6 & = & -\frac{1}{6} a^6 s^3 \left(1 + 3 d - \frac{216}{35} \frac{c}{\sqrt{s}} \right) 
\\ 
A_8 & = & \ldots 
\label{eq_A2pfirst}
\end{eqnarray}
More generally, one finds
\begin{equation}
A_{2p} = \frac{(-1)^p}{p!} (sa^2)^p \left( 1 +
p d - \frac{3 (2^p p! p)^2}{2 (2p+1)!} \frac{c}{\sqrt{s}} 
\right) 
\label{eq_A2pgeneral}
\end{equation}
From this result and using Eq.~(\ref{eq_r2pA2p}) one immediately verifies that
the moments of the distribution are given by the Eqs.~(\ref{eq_bepert}) and 
(\ref{eq_Rppert}). Using Eq.~(\ref{eq_alphapAp}) one justifies similarly 
Eq.~(\ref{eq_alphap}) for the non-Gaussian parameter $\alpha_p$. 

These moments completely determine the segmental distribution $G(r,s)$ which is 
indicated in Eq.~(\ref{eq_Grspert}). While at least in principle this may be done 
directly by inverse Fourier-Laplace transformation of the correction $\delta G(q,t)$ 
to the generating function it is helpful to simplify further Eq.~(\ref{eq_deltaGqt}).
We observe first that $\delta G(q,t)$ does diverge for strictly incompressible systems 
($v\rightarrow \infty$) and one must keep $v$ finite in the effective 
potential whenever necessary to ensure convergence (actually everywhere but in the 
diagram corresponding to the interaction between two external monomers). Since
we are not interested in the wave vectors larger than $1/\xi$ we expand 
$\delta G(q,t)$ for $\xi\rightarrow 0$ which leads to the much simpler expression
\begin{equation}
\delta G (q,t) 
\approx -\frac{v\xi q^2}{3 \pi a^2(a^2q^2+t)^2} + \frac{v \xi^2}{4\pi a^6}\frac{a\sqrt
{t}(3 a^2q^2+t)}{(a^2q^2+t)^2} - \frac{v \xi^2}{4\pi a^6}\frac{\mbox{Arctan}[\frac{a q}{
\sqrt{t}}]}{q} + o(v\xi^3).
\label{eq_Gdl}
\end{equation}
The first term diverges as $\sqrt v$ for diverging $v$. It renormalizes the 
effective bond length in the zero order term which is indicated in the first
line of Eq.~(\ref{eq_Grspert}).
%
%
The next two terms scale both as $v^0$. Subsequent terms must all vanish for diverging $v$ 
and can be discarded. It is then easy to perform an inverse Fourier-Laplace transformation
of the two relevant $v^0$ terms. This yields
\begin{equation}
\delta G(x,s) =
G_0(x,s) \frac{c}{\sqrt s}
\frac{3\sqrt{\pi}}{4} \left(-\frac{2}{x}+\frac{3x}{2}
- \frac{x^3}{8}\right)
\label{eq_Greal}
\end{equation}
with $x=r/a \sqrt{s} = \sqrt{6} n$. This is consistent with the expression given in
the second line of Eq.~(\ref{eq_Grspert}). 

We note finally that the intramolecular form factor 
$F(q) = \frac{1}{N} \sum_{n,m=1}^N \la \exp(i {\bf q} \cdot ({\bf r}_n - {\bf r}_m) \ra$ 
of asymptotically long chains can be readily obtained from Eq.~(\ref{eq_Gdl}). 
Observing that $\la \exp(i {\bf q} \cdot ({\bf r}_n - {\bf r}_m) \ra =
\int d^3{\bf r} \exp(i {\bf q}\cdot {\bf r}) G(r,s) = G(q,s)$
one finds
\begin{equation}
\delta F(q) = 2 \int ds \ \delta G(q,s) = 2 \ \delta G(q,t=0) =
- 2 \frac{v \xi^2}{4 \pi a^6} \frac{\pi/2}{q},
\label{eq_deltaFq}
\end{equation}
where we used the third term of Eq.~(\ref{eq_Gdl}) in the last step.
The first term in Eq.~(\ref{eq_Gdl}) is discarded as before, since it renormalizes
the effective bond length in the reference form factor:
$F_0(q) = 12/b^2 q^2 \Rightarrow 12/\be^2 q^2$.
It follows, hence, that within first-order perturbation theory
\begin{equation}
F(q) = F_0(q) + \delta F(q) 
\approx F_0(q) \left( 1 - \frac{3}{8} \frac{\be q}{\be^3\rho} \right)
\end{equation}
as indicated by Eq.~(\ref{eq_Fq}) in the Conclusion. This is equivalent to
the result
$1/F(q) - 1/F_0(q) \approx q^3/32\rho$
discussed in Refs.~\cite{WBJSOMB07,BJSOBW07} for polymer melts
and anticipated by Sch\"afer \cite{SchaferBook} by renormalization
group calculations of semidilute solutions.
%
%

\newpage
\begin{table}[t]
\begin{tabular}{|l|l||c|c|c|c|c|c|c|}
\hline
$N$  & $ $\nch$  $ & $\taue$   & $\Rend$& $\Rgyr$&$\be(N)$ 
& $\sqrt{6} \bg(N)$ \\ \hline
16   & $2^{16}$ &  $1214$        & 11.7 & 4.8  & 2.998   & 2.939 \\
32   & $2^{15}$ &  $3485$        & 17.1 & 7.0  & 3.066   & 3.030 \\
64   & $2^{14}$ & $1.1\cdot10^4$ & 24.8 & 10.1 & 3.116   & 3.094 \\
128  & 8192     & $3.3\cdot10^4$ & 35.6 & 14.5 & 3.153   & 3.139 \\
256  & 4096     & $1.0\cdot10^5$ & 50.8 & 20.7 & 3.179   & 3.171 \\
512  & 2048     & $3.2\cdot10^5$ & 72.2 & 29.5 & 3.200   & 3.193 \\
1024 & 1024     & $1.0\cdot10^6$ & 103  & 42.0 & 3.216   & 3.212 \\
2048 & 512      & $3.2\cdot10^6$ & 146  & 59.5 & 3.227   & 3.223 \\
4096 & 256      & $9.7\cdot10^6$ & 207  & 85.0 & 3.235   & 3.253 \\
8192 & 128      & $2.9\cdot10^7$ & 294  & 120  & 3.249   & 3.248 \\
\hline
\end{tabular}
\vspace*{0.5cm}
\caption[]{Various static properties of dense BFM melts of 
number density $\rho = 0.5/8$:
the chain length $N$, the number of chains $\nch$ per box, 
the relaxation time $\taue$ characterized by the diffusion of the monomers
over the end-to-end distance and corresponding to the circles indicated in Fig.~\ref{fig_taue}, 
the root-mean-squared chain end-to-end distance $\Rend$ and 
the radius of gyration $\Rgyr$ of the total chain ($s=N-1$).
The last two columns give estimates for the effective bond length
from the end-to-end distance, $\be(N) \equiv \Rend/(N-1)^{1/2}$, and
the radius of gyration, $\bg(N) \equiv \Rgyr / \sqrt{N}$.
The dashed line in Fig.~\ref{fig_Rslog} indicates $\be(N)^2$.
Apparently, both estimates increase monotonicly with $N$ 
reaching $\be(N) \approx \sqrt{6} \bg(N) \approx 3.2$ for the largest chains available. 
Note that $\sqrt{6} \bg(N) < \be(N)$ for smaller $N$.
%
%
\label{tab_BFMmelt}}
\end{table}

\newpage
\begin{table}[t]
\begin{tabular}{|l||c|c|c|}
\hline
Property                                 & BFM              & BSM           \\ \hline
Length unit                              & lattice constant & bead diameter \\
Temperature $\kB T$                      & 1                & 1             \\
Number density $\rho$                    & 0.5/8            & 0.84          \\
Linear box size $L$                      & 256              & $\le 62$      \\
Number of monomers $\nmon$               & 1048576          & $\le 196608$  \\
Largest chain length $N$                 & 8192             & 1024          \\ \hline
Mean bond length $\langle |{\bf l}_n| \rangle$    & 2.604   & 0.97          \\
$ l = \langle {\bf l}_n^2 \rangle^{1/2}$ & 2.636            & 0.97          \\
Effective bond length $\be$              & 3.244            & 1.34          \\
$\rho \be^3$                             & 2.13             & 2.02          \\
$C_{\infty} = (\be/l)^2$                 & 1.52             & 1.91          \\
$\lp = l (C_{\infty} +1 )/2$             & 3.32             & 1.41          \\         
$\ce \equiv \sqrt{24/\pi^3}/\rho \be^3 $ & 0.41		    & 0.44          \\
$c_1/\ce $                               & $1.0$            & $1.2$ \\
$c_2/\ce $                               & $1.0$            & $1.1$ \\
$c_3/\ce $                               & $1.0$            & $1.0$ \\
$c_4/\ce $                               & $1.1$            & $1.2$ \\
$c_5/\ce $                               & $1.1$            & $0.9$ \\
$\cP = c_1 (\be/l)^2/8$                  & 0.078            & 0.124         \\
Dimensionless compressibility $g$        & 0.245            & 0.08          \\
Compression modulus $v \equiv 1/g\rho$         & 66.7             & 14.9          \\
$\Gi \equiv \sqrt{v \rho} / \be^3 \rho$       & 0.96             & 1.8           \\
\hline
\end{tabular}
\vspace*{0.5cm}
\caption[]{Comparison of some static properties of dense BFM and BSM melts.
The first six rows indicate conventions and operational parameters.
%
%
The effective bond length $\be$ and the swelling coefficients $\cp$ 
(defined in Eq.~(\ref{eq_Rs_key})) are determined from the first
five even moments of the segmental size distribution. 
The dimensionless compressibility $g = S(q\rightarrow 0)/\rho$ has been obtained 
from the total static structure factor 
$S(q) = \frac{1}{L^3} \sum_{k,l=1}^{\nmon} \la \exp(i {\bf q} \cdot ({\bf r}_{k} - {\bf r}_{l})) \ra$
in the zero wave vector limit as shown at the end of Ref.~\cite{BJSOBW07}.
The values indicated correspond to the asymptotic long chain behavior.
Properties of very small chains deviate slightly.
\label{tab_BFMBSMmelt}}
\end{table}
%
%
%
%
%
%
%

\newpage
\clearpage

\newpage
\clearpage

\newpage
\clearpage

\newpage
\clearpage

\newpage
\clearpage

\newpage
\clearpage

\newpage
\clearpage
\newpage
\clearpage

\newpage
\clearpage

\newpage
\clearpage
\newpage
\clearpage
\newpage
\newpage
\clearpage


\end{document}